\documentclass[journal]{IEEEtran}
\IEEEoverridecommandlockouts
\usepackage{cite}
\usepackage{amsmath,amssymb,amsfonts}
\usepackage{algorithmic}
\usepackage{graphicx}
\usepackage{textcomp}
\usepackage{xcolor}
\usepackage{siunitx}
\usepackage{booktabs} 
\usepackage{multirow}
\usepackage{booktabs,threeparttable}
\usepackage{cleveref}
\usepackage[nolist,nohyperlinks]{acronym}
\usepackage{makecell}
\ifCLASSOPTIONcompsoc
    \usepackage[caption=false, font=normalsize, labelfont=sf, textfont=sf]{subfig}
\else
\usepackage[caption=false, font=footnotesize]{subfig}
\fi

\usepackage{siunitx}
\def\BibTeX{{\rm B\kern-.05em{\sc i\kern-.025em b}\kern-.08em
    T\kern-.1667em\lower.7ex\hbox{E}\kern-.125emX}}
\begin{document}

\title{Characterization of Spatial-Temporal Channel Statistics from   Measurement Data at D Band\\}

\author{Chathuri~Weragama,~\IEEEmembership{Student Member, IEEE,}
        Joonas~Kokkoniemi,~\IEEEmembership{Member, IEEE,}
        Mar~Francis~De~Guzman,~\IEEEmembership{Member, IEEE,}
        Katsuyuki~Haneda,~\IEEEmembership{Member, IEEE,}
        Pekka~Ky\"osti,
        ~and~Markku~Juntti~\IEEEmembership{Fellow, IEEE.}
\thanks{This work was supported by the European Union Smart Networks and Services Joint Undertaking (SNS JU) under grant agreement no. 101096949 (TERA6G). This research was also supported by the Research Council of Finland (former Academy of Finland) 6G Flagship Programme (Grant Number: 346208).
 }
\thanks{Chathuri Weragama, Joonas Kokkoniemi, Pekka Ky\"osti, and Markku Juntti are with Centre for Wireless Communications (CWC), University of Oulu, 90014 Oulu, Finland (email: chathuri.weragama@oulu.fi, joonas.kokkoniemi@oulu.fi, pekka.kyosti@oulu.fi, markku.juntti@oulu.fi) }
\thanks{Mar~Francis~De~Guzman and Katsuyuki~Haneda are with Department of
Electronics and Nanoengineering, Aalto University School of Electrical Engineering, 02150 Espoo, Finland (email: francis.deguzman@aalto.fi, katsuyuki.haneda@aalto.fi)  }

}

\begin{acronym}
  \acro{LOS}{line of sight}
  \acro{NLOS}{non-line of sight}
  \acro{NoM}{number of measurements}
  \acro{NoD}{number of data points}
  \acro{mmWave}{millimeter-Wave}
  \acro{6G}{sixth generation}
  \acro{5G}{fifth generation}
  \acro{URLLC}{ultra reliable low latency communications}
  \acro{mMTC}{massive machine type communications}
  \acro{eMBB}{enhanced mobile broadband}
  \acro{MPC}{multi-path component}
  \acro{MIMO}{multiple-input multiple-output}
  \acro{Tx}{transmitter}
  \acro{Rx}{receiver}
  \acro{RF}{radio frequency}
  \acro{CI}{close-in}
  \acro{FI}{floating intercept}
  \acro{FSPL}{free space path loss}
  \acro{KS}{Kolmogorov- Smirnov}
  \acro{KL}{Kullback-Leiber}
  \acro{AD}{Anderson-Darling}
  \acro{CDF}{cumulative distribution function}
  \acro{FT}{Fourier transform}
  \acro{PDP}{power delay profile}
  \acro{NPD}{normalized power distribution}
  \acro{NDD}{normalized delay distribution}
  \acro{NoP}{number of paths}
  \acro{AoA}{angle of arrival}
  \acro{AoD}{angle of depature}
  \acro{GBSM}{geometry-based stochastic channel model}
  \acro{NGSM}{nongeometrical stochastic channel model}
  \acro{S-V}{Saleh-Valenzuela}
  \acro{SISO}{single-input single-output}
  \acro{ULA}{uniform linear array}
  \acro{MED}{maximum excess delay}
  \acro{PDF}{probability distribution function}
  \end{acronym}

\maketitle

\begin{abstract}
Millimeter-Wave (mmWave) (30-300 GHz) and D band (110–170 GHz) frequencies are poised to play a pivotal role in the advancement of sixth-generation (6G) systems and beyond with increased demand for greater bandwidth and capacity. This paper focuses on deriving a generalized channel impulse response for mmWave communications, considering both outdoor and indoor locations for line-of-sight (LOS) and non-line-of-sight (NLOS) scenarios. The analysis is based on statistical insights obtained from measurements conducted at distinct locations with a center frequency of 142 GHz, examining parameters such as path gain, delay, number of paths (NoP), and angle distributions. Whereas different distributions serve as candidate models for the gain of LOS communications, only specific distributions accurately describe the NLOS gain, LOS and NLOS delay, LOS and NLOS NoP, and LOS and NLOS angular distributions. The channel is modeled based on geometry-based stochastic channel modeling (GBSM) with parameters derived from the statistical analysis. The maximum excess delay is used as a metric to evaluate the performance of the proposed model against empirical data. 

\end{abstract}


\section{Introduction}

\IEEEPARstart{I}{n} the past decade, an exploration into \ac{mmWave} frequencies up to 100 GHz has paved the way for their integration into \ac{5G} communication systems, offering improved communications, localization, and sensing catering to specific requirements within the industry \cite{wymeersch2021integration}. However, the evolving demands of \ac{6G} communication, marked by higher data rates, seamless connectivity, and greater device density to support \ac{URLLC}, \ac{mMTC}, and \ac{eMBB}, necessitate a deeper investigation into propagation and channel characteristics at even higher frequencies. 
The D band, spanning from 110 GHz to 170 GHz, has emerged as a focal point for addressing these requirements in 6G. Several studies have been carried out to understand the material characteristics and channel modeling at D band \cite{9611219, wang2022reflection,9815763,9769465} considering reflection, scattering and diffraction phenomena at this frequency range. Whereas existing research has endeavored to characterize path loss, angular spread, and inter-cluster characteristics at the D band using indoor measurements \cite{7095539,8936412}, challenges on channel modeling persist due to frequency-selective molecular absorption, high attenuation, reduced penetration through solid objects, and diffuse scattering \cite{petrov2016terahertz}. The short wavelengths in the D band cause increased scattering from objects that are considered smooth at lower frequencies \cite{7499284}, further increasing the complexity of multipath modeling for \ac{mmWave} communications.

One versatile technique for channel characterization is stochastic modeling, as it can capture the statistical behavior of wireless channels for different scenarios \cite{han2022terahertz}. An extension of stochastic channel models, \ac{GBSM} combines geometric and stochastic methods to formulate a channel model. By considering the relative geometries of link-ends and wave interaction objects, \ac{GBSM} enables more generalized channel models for different environments operating in the same frequency compared to purely stochastic models. This contrasts with two other methods: deterministic channel modeling \cite{yang1998ray, son1999deterministic}, which focuses on specific environments, and hybrid channel modeling \cite{lecci2020quasi,maltsev2014quasi, zhu20213gpp}, which combines deterministic and stochastic approaches. \ac{S-V} model\cite{saleh1987statistical}, a stochastic channel model, has been the basis for development of \ac{GBSM} for frequencies below 100 GHz communications \cite{gustafson2013mm,park1998analysis}, enabling more generalized channel models. A channel model based on \ac{GBSM} for a factory environment at 140 GHz has been derived in \cite{10637278}, we intend to derive a channel model which is applicable for a variety of different environments such as airports, shopping malls and open halls with the intention of providing a basic channel model for signal processing algorithm development. 
 To address this, we extend our preliminary work\cite{weragama2024characterization} on indoor channel propagation characteristics in D band communications to provide a comprehensive analysis on \ac{mmWave} communication channels for frequencies above 100 GHz. Our previous study focused on path gain, delay, and \ac{NoP} for indoor environments. In this paper, we significantly expand the scope and depth of our investigation.

The major contributions of this study are three-fold to:

\begin{enumerate}
    \item Present a detailed characterization of \ac{mmWave} radio propagation for both indoor and outdoor locations. This analysis reveals novel insights into the behavior of path gain, delay, and \ac{NoP} across diverse environments, providing valuable data for the design of future \ac{mmWave} systems.
    \item Develop a channel impulse response for \ac{MIMO} \ac{mmWave} communications based on the \ac{GBSM}. This model offers a versatile tool for researchers and engineers to accurately simulate and predict \ac{mmWave} channel behavior in various scenarios.
    \item Identify and discuss the specific benefits and limitations of deriving channel statistics solely from measurement data. This contribution provides crucial insights for researchers, highlighting areas where measurement-based approaches excel and where they may fall short, thus guiding future research methodologies in \ac{mmWave} channel modeling.
\end{enumerate}

Insights into statistical parametrization of  \ac{mmWave} channels are obtained based on measurements carried out in different indoor and outdoor locations. We characterize the path gain, delay and \ac{NoP} parameters for indoor and outdoor radio links under \ac{LOS} and \ac{NLOS} conditions. A detailed analysis on \ac{AoA}/ \ac{AoD} was conducted in \cite{de2024comparison} for the same data set. The mentioned channel parameters related to the collected data undergo comparisons with theoretical distributions to elucidate their inherent properties. We aim to aggregate all measured paths observed from specific locations to derive statistical distributions for amplitude or path gain, delay, and the \ac{NoP}.  The distributions, combined with the distribution of angular spread, a predetermined antenna array configuration, and \ac{FSPL} will make it possible to generate random wideband \ac{MIMO} channel by arbitrarily summing paths from the derived distributions.

This paper is organized as follows. Section \ref{sec:channel model} provides a description of the channel model we will be using in our study and Section \ref{sec:mesaure} provides an overview of the channel sounding procedure, while Section \ref{sec:methodology} delineates the methodology employed for post-data processing of the measured channels and analysis of statistical parameters. Section \ref{sec:results} presents the findings of our analysis and associated statistical metrics, while Section \ref{sec:verfi} provides verification and discussion on the results of this analysis. Finally, Section \ref{sec:conclusion} encapsulates the conclusions drawn from our study.

\section{Geometric Stochastic Channel Model}\label{sec:channel model}

Channel models can be determined for several purposes with various granularities. The simplest case is for predicting the overall signal attenuation, i.e., path loss modelling \cite{Sooyoung2016mmWave, Youngbin2014dual, Tataria2020channel, Haneda2016}.  These are based on fitting a path loss curve over the measurement data and using a random variable to model fading/shadowing. For other needs, delay, spatial, and Doppler domains can also be considered. For MIMO modelling, typically channel impulse responses or transfer functions are to be predicted per \ac{Tx}/\ac{Rx} antenna port pair. Path loss modelling using omni-directional assumption is not applicable at the considered radio frequencies, since high antenna gains are a necessity on any practical systems, leading to highly directive antenna systems, i.e., strongly non-uniform antenna patterns.
\subsection{Channel Model}
We aim at deriving the statistics of the channels by deriving impulse response or transfer functions per \ac{Tx}/\ac{Rx} pair combining the antenna array patterns, where we analyse the characteristics of each \ac{MPC} of the channel in order to estimate the channel statistics. Our ultimate goal is to derive stochastic \ac{MIMO} channel models by utilizing the statistics obtained from the measurements. While a similar approach to \ac{MIMO} channel modeling has been undertaken in \cite{Das2024mimo}, it is noteworthy that our statistics are derived solely from measurement data.
The \ac{GBSM} for \ac{mmWave} communications presented in this study is an extension of the \ac{S-V} channel model, but without clustering, which differs to the channel model used in\cite{10637278}. This channel model can be expressed mathematically as:
\begin{equation}\label{equ: channel model}
    \mathbf{\text{H}} = \sum^{L}_{l=1}\delta^F_l(f,\tau_l)\delta^A_l\mathbf{a}_{\text{r}}(\phi_l){\mathbf{a}}_{\text{t}}^\text{H}(\theta_l)e^{-i (2\pi f\tau_l+\beta)},
\end{equation}
where $L$ is the number of paths, $\delta^F_l$ is the free-space propagation gain of the $l^{th}$ path, which depends on the random delay ($\tau_l$) of the path. $\delta^A_l$ is the excess path gain of the $l^{th}$ path, which is dependent on the environment, $\mathbf{{a_{t/r}}}=A_o[1 \text{ }  e^{-j 2 \pi\sin(\theta/\phi+\theta_l/\phi_l)} \ldots e^{-j 2 \pi(N_{T/R}-1)\sin(\theta/\phi+\theta_l/\phi_l)}]^T $ is the \ac{ULA} response  vectors, where $N_{T/R}$ is the number of antennas, $A_o$ is a complex amplitude representing the antenna gain, which in a general case depends on the antenna radiation pattern and it is a vector if the antenna amplitudes are not equal or are adjustable. The departure and arrival angles of a path seen from the phase center of the antenna array are represented by $\theta_l$ and $\phi_l$. It is worth noting that this antenna array response will be a scalar when a \ac{SISO} scenario is considered. The last exponential term represents the phase shift experienced by the signal. For narrowband transmissions, this phase can be treated as entirely random for each path, since the frequency-dependent part $(2\pi f\tau_l)$ doesn't vary much across the small bandwidth, and it combines with the random phase offset $\beta$. However, for wideband channels, we must explicitly consider the frequency dependency of this term $(2\pi f\tau_l)$ across the larger bandwidth, as it leads to different phase shifts for different frequency components of the signal. The $\beta$ term may still be random, but the overall phase  $(2\pi f\tau_l+\beta)$ behavior is dominated by the frequency-dependent term in wideband scenarios. A similar channel model has been used in recent signal processing research \cite{10286447,10214237}.

\subsection{Parameters of the Model}

A proper understanding on the behaviour of path gain, delay, \ac{NoP}, and the \ac{AoA}/ \ac{AoD} are required to derive a channel model based on (\ref{equ: channel model}). Once the \ac{AoA} and \ac{AoD} are determined, the antenna array response can be predetermined based on the antenna type. \ac{FSPL} can be calculated with the delay of the paths known. 
We aim to determine the statistical behaviour of these parameters related to \ac{mmWave} communication based on measurement data with the intention of deriving a channel matrix for \ac{mmWave} communication based on (\ref{equ: channel model}). An analysis on path gain, delay and \ac{NoP} is performed in this study to understand the distribution of these parameters, as a detailed analysis on \ac{AoA}/ \ac{AoD} is performed in \cite{de2024comparison} for the same data set concluding these parameters follow a log-normal distribution. Once the distributions of the aforementioned paramters are identified we select random values from these distributions to build the channel model presented in (\ref{equ: channel model}).

It is important to note that the \ac{GBSM} model presented in this study is based solely on measurement data, which introduces certain limitations. For instance, due to the finite number of available measured radio links, it is challenging to analyze whether the path gain and delay can be represented as a joint distribution. This analysis would require the very large number of link measurements that are difficult to obtain by the setup of sub-THz channel sounder in this paper. Potential solution would be to calibrate ray tracing simulators with real measurement data and then use simulated data to inflate the dataset and therefore get more data for the joint distribution analysis. However, this is out of scope of this work.
Furthermore, the limited number of radio link measurements constrained our ability to draw definitive conclusions for distributions of some parameters, such as \ac{NoP}. These limitations highlight the challenges inherent in developing comprehensive channel models based on empirical data and underscore the need for ongoing research and data collection in this field.

In the following section, we present a comprehensive set of channel measurements conducted across various indoor and outdoor locations. These measurements will allow us to examine the practical phenomenon of the channel characteristics outlined in our theoretical model, including path gain, delay, and the \ac{NoP}. 

\section{Channel Measurement Data } \label{sec:mesaure}

The channel measurements on the D band were conducted by Aalto university in various locations in Helsinki region, Finland \cite{DeGuzman2022entrance, nguyen2018shopping, nguyen2021airport}. Several datasets from these measurements are openly available in \cite{de_guzman_2023_7640353}. The chosen indoor locations comprise commercial buildings: Sello shopping mall in the city of Espoo, Helsinki-Vantaa Airport, and the entrance hall of Aalto University in Espoo. Notably, measurements at the Aalto University entrance hall were collected during two separate periods, and treated as distinct locations for analysis purposes. Outdoor locations consist of Aalto University campus area, Helsinki city center area and Espoo city residential area. 

All indoor locations have a relatively open floor plan with also some notable differences. The shopping mall is a multistory building with open areas but also more closed ones. The airport is mostly a single floor, but very open, and the Aalto university location is partially open space, but with more obstacles and features. When outdoor locations are considered, the campus area can be considered as a suburban area where walls of the buildings in the
area are mainly made up of bricks and have glass windows and doors with metallic frames. The Helsinki city center area is surrounded by commercial buildings forming a street canyon. The residential area is similar to the city center area but less dense with buildings. A detailed descriptions of these locations are provided in \cite{selimis2022path}. However, unlike the indoor locations the outdoor location measurements were conducted while moving objects were present in these locations.

The exact measurement setups have been detailed in various previous works, such as in \cite{DeGuzman2022entrance} for the Aalto University entrance hall, \cite{nguyen2018shopping} for the shopping mall, and \cite{nguyen2021airport} for the airport and shopping mall. 
The shopping mall and airport measurements were conducted at 143.1 GHz center frequency, while the remaining ones were conducted in the 142 GHz frequency with 4 GHz bandwidth. The \ac{Tx} end was biconical omnidirectional antenna (0 dBi) and the \ac{Rx} end was a horn antenna on rotating platform (19 dBi). The \ac{RF} power was about \qty{-7}{dBm} and the observed noise floor was $-128$ dBm, with an assumed margin of 10dB. The measurement details related to these seven locations are provided in Table \ref{tab:measurement_campaign}. In this table Sello, Airport, TUAS, and TUAS2 locations refer to Sello shopping mall,  Helsinki-Vantaa Airport, initial measurement campaign at the entrance hall of Aalto University, and the second measurements for indoor locations, respectively. While campus, city, and residential locations refer to Aalto University area, Helsinki city-center area and the Espoo-city residential area for outdoor locations.

\begin{table*}[]
 \caption{Details on measurement campaign}
    \label{tab:measurement_campaign}
    \centering
    \begin{tabular}{c|c|c|c|c|c|c|c}\hline
        Measurement Details & Sello & Airport & TUAS& TUAS2 & Campus & City & Residential  \\ \hline
        RF(GHz)& 141.5-145.1 & 141.5-145.1 & 140-144& 140-144 & 140-144 & 140-144 & 140-144 \\ \hline
        Tx Antenna Height(m) & 1.89 & \makecell{1.7 above
        \\ 2nd floor} & 1.85& 1.85 & 1.85 & 1.85 & 2.00 \\ \hline
        Rx Antenna Height(m) & 1.89 & \makecell{2.1 above\\ 3rd floor} & 1.85 & 1.85  & 1.85 & 1.85  & 1.85   \\ \hline
        EIRP(dBm) & -12 & -12 & 5& 5 & 5 & 5& 5 \\ \hline
        Rx Azimuth Range($^o$) & 0-360 & 0-12, 245-360 & 40-250& \makecell{-90-180(Rx1)\\40-250(Rx2) \\ 110-290(Rx3)} & 0-355 & Mostly 0-355 &  Mostly 0-355  \\ \hline
        Azimuth Step($^o$) & 6 & 5 & 10& 5 &  5  &  5  &  5  \\ \hline
        Environment Type & Indoor & Indoor & Indoor& Indoor& Outdoor & Outdoor& Outdoor  \\ \hline
        Link Distance Range(m) & 3-65 & 15-51 & 3-47& 3-66 & 2-172 & 20-175 & 10-178  \\ \hline

    \end{tabular}
   
\end{table*}


The same measurement data has been analysed before with respect to the fading statistics \cite{Papasotiriou2021fading }. Our work here differs from that by aggregating the data per location in order to obtain statistically independent distributions (within the limits of the amount of measurement data available) as any single measurement link is strongly related to the environment geometry. When aggregating all the measurement positions per measurement location, we will obtain a stochastic data set as possible given the limitations of the real world measurements. The process of obtaining the statistics and channel modeling is described in the next section.

\section{Modelling Methodology} \label{sec:methodology} 

\subsection{Data Manipulation}

We focus on analyzing the \ac{PDP}, path gain distribution, delay distribution, and the \ac{NoP} for both \ac{LOS} and \ac{NLOS} scenarios with the objective of modeling a channel frequency response for the D band communication. An analysis of angular spread of the \ac{MPC}s has been carried out for the same data in previous studies \cite{de2024comparison}, which suggests that it follows a log-normal distribution. 

The data collection process discussed in Section \ref{sec:mesaure} involved aggregating all \ac{MPC}s observed from radio channel measurements in the measurement location into a unified dataset. Following this approach, the channel data underwent normalization of the \ac{FSPL} for the gain estimates of MPCs  based on the estimated delay of the same MPCs. This then mitigates the influence of distance and allows to examine power and delay distributions of MPCs. This way the free space expansion of the waves is neglected and we can extract the loss related only to wave-object interaction. That is, we obtain the loss due to reflections, scattering, diffraction, multiple bounces, etc. to calculate the excess propagation loss. From this, we can calculate the statistics of the excess loss and add the free space loss back later when the path gain of MPCs is needed. Hence, this method aims to treat the far away paths equally to derive more accurate statistics for the excess loss. 

The normalized received path gain, compensating for \ac{FSPL} and obtain only the impact of the channel on the received power, is expressed as
\begin{equation}\label{equ:power_norm}
P_{\mathrm{n}(\si{dB})} = P_{(\si{dBm})} - 20\log_{10}\left(\frac{4\pi f d}{c}\right),
\end{equation}
where $P_{\mathrm{n(dB)}}$ denotes the normalized power, $P_{\mathrm{(dBm)}}$ is the measured power, and $d$, $f$, and $c$ represent the length of the path, operating frequency, and speed of light in free space, respectively. The delay, on the other hand, is normalized with the delay of the first arrival to model the delay spread with the distributions. The normalization of delay was calculated as
\begin{equation}
\label{equ: delay_norm}
\tau_{n} = \tau_{l} - \tau_{1},
\end{equation}
where $\tau_{n}$, $\tau_{l}$, and $\tau_{1}$ correspond to the normalized delay, measured delay of the $l^{th}$ MPC, and the delay of the first-arrived path, respectively. Modeling the \ac{NoP} poses challenges due to inherent sparsity in the measurements. However, despite these challenges, we were able to derive reasonably accurate statistics for the \ac{NoP} as well. This normalized power and delay data elucidate the behavior of power and delay profiles within the D band.  

\subsection{Analysis of Distributions}

This analysis mainly focuses on deriving statistical distributions to represent path gain, delay, and \ac{NoP} based on measured data. Throughout this analysis, we treat path gain and delay distributions independently of one another due to two main reasons. 
\begin{itemize}
    \item High losses prevent signals from bouncing indefinitely, hence the power and delay are less connected than in low frequency bands where large numbers of interactions lead to longer paths.
    \item The amount of data is too low to perform reliable joint distribution statistics. Producing such high amounts of data is very challenging at these frequencies, because omnidirectional measurements are in practice not possible and directional measurements are always slower to perform.
\end{itemize}

To evaluate the statistical distributions derived from our measured data, we employed several analytical techniques. There are three main approaches to analyse the adequacy between the empirical and theoretical distributions, namely graphical methods, formal tests and normality tests. However,  evaluating the goodness of  fit of a distribution based on single evaluation metrics will lead to inaccurate results. Therefore, we have used both graphical and normality tests for our  evaluation. When the normality tests are considered, we have evaluated  many normality tests, including  \ac{KS} test \cite{massey1951kolmogorov}, the \ac{KL} divergence \cite{kullback1951information}, Shapiro-Wilk, Lilliefors and the \ac{AD} test. Though the literature suggests\cite{yap2011power} that the \ac{AD} test has a superior performance in comparison to \ac{KS} test, since there are a number of distributions under consideration, \ac{KS} test was used as the normality  test  due to its applicability in various distributions. The \ac{KS}  test statistic for a given $m$ ordered data points is defined as 
\begin{equation}
    T = \sup_x|F_{m}(x)-F(x)|,
\end{equation} where $\sup_x$ is the supremum of the set of distances, $F_{m}(x)$ is the empirical \ac{CDF} and $F(x)$ is the \ac{CDF} of the theoretical distributions. A p-value is calculated based on \ac{KS}  statistic and if the p-value exceed the critical value with respect to its significant value it is considered that the theoretical distribution is a good fit for the empirical distribution and vise versa. For visual representations, we have used both Q-Q plots and visualisation of \ac{CDF}s (\Crefrange{fig:NPD - LOS}{fig : NoP-NLOS-fitting-outdoor}) of theoretical and empirical distributions for our evaluation. Q-Q plots compare the quantiles of the first dataset against those of the second dataset. A perfect match between datasets from the same distribution would result in a plot following a $y=x$ line with a correlation coefficient of 1. We have used the correlation coefficient of this plot to serve as a metric for assessing the congruence between the two distributions.

\subsection{Maximum Excess Delay}\label{sub:delay_spread}

The \ac{MED} can be used as a metric to evaluate the performance of our model against the empirical data. The \ac{MED} of a channel can be calculated as 
\begin{equation}\label{equ:delay_spread}
    \tau = \tau_\text{L} -\tau_1,
\end{equation}
where $\tau_\text{L}$ and $\tau_1$ represents the delay of the last arriving \ac{MPC} and the delay of the first arriving \ac{MPC} respectively.  To calculate the \ac{MED} using the model as mentioned above, we have randomly drawn values for delay and gain from the respective distributions, which we have analysed for each scenario, and then calculated the power for each \ac{MPC} by adjusting the gain for \ac{FSPL} which was calculated based on the delay. Once the power values have been calculated we have drawn the delay components of the \ac{MPC}s which have power values above the noise floor of the measurements. Once these delay values have been found  we calculate the \ac{MED} of the \ac{MPC}s according to  (\ref{equ:delay_spread}). Then these values were compared to the empirical \ac{MED} for each location.
The analysis of this study has been conducted in a Python environment using libraries numpy \cite{harris2020array}, pandas\cite{mckinney2010data}, and scipy.stats \cite{2020SciPy-NMeth} for data manipulation, data organization and statistical fitting, respectively. The main tools that aided the visual representations of the said analysis are the matplotlib\cite{hunter2007matplotlib} and the seaborn\cite{waskom2021seaborn} libraries. It is worth noting that statistical parameters related to distribution fitting in this study are compatible with Python scipy.stats library.

\section{Statistical Channel Parameters }\label{sec:results}

As mentioned in Section \ref{sec:methodology}, the data was collected for both indoor and outdoor locations covering \ac{LOS} and \ac{NLOS} propagation scenarios. This section discuss the results of the analysis conducted on these locations together with the statistics related to the analysis. The \ac{NoM}, and the \ac{NoD} for each location is presented in Table \ref{tab:Number of Data Points}. As discussed in the previous section,  Sello, Airport, TUAS, and TUAS2 locations refer to Sello shopping mall, Helsinki-Vantaa Airport, initial measurement campaign at the entrance hall of Aalto University and the second measurements respectively for the indoor locations and for outdoor locations Campus, City, and Residential referees to Aalto University outdoor area, Helsinki city area and   residential area of Espoo. We have analysed the measurements of \ac{MPC}s based on location to understand the behaviour of power, delay, and \ac{NoP} when propagating at D band.

 \begin{table}[t]
\caption{Number of data points per location}

    \label{tab:Number of Data Points}
    \centering
\begin{tabular}{c |l| c c|c c} \hline
    \multirow{2}{*}{Location Type }& \multirow{2}{*}{Location } &\multicolumn{2}{c}{LOS} & 
        \multicolumn{2}{c}{NLOS} \\ 
        & & NoD & NoM & NoD & NoM \\ \hline
        \multirow{4}{*}{Indoor}& Sello&304 & 16 & 29 &2 \\ 
        & Airport & 375&10 & 41 & 1\\ 
        & TUAS & 29 &5 & 378 & 21\\ 
        & TUAS2 & 268 &16 & 1812& 61\\ \hline
    \multirow{3}{*}{Outdoor}& Campus&924 & 35 & 14 & 8\\ 
    & City & 768&12 & 435 & 21\\
    & Residential & 477 &21 & 183 & 27\\ \hline

    \end{tabular}

\end{table}

\subsection{Power Delay Profile }
\begin{figure}[t]
    \centering
    \subfloat[Sello]{%
       \includegraphics[scale=0.5]{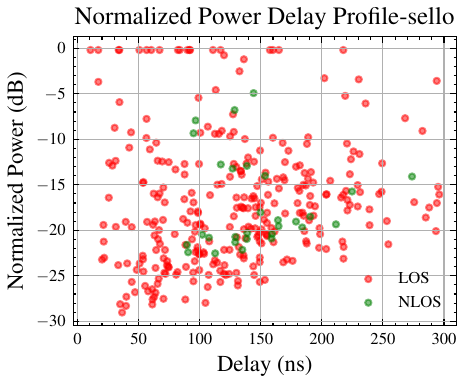}}
    \subfloat[Airport]{%
        \includegraphics[scale=0.5]{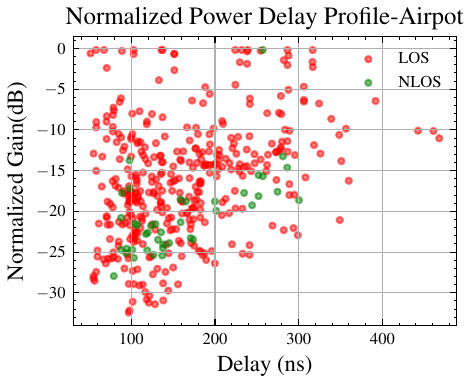}}
    \\
    \subfloat[TUAS]{%
        \includegraphics[scale=0.5]{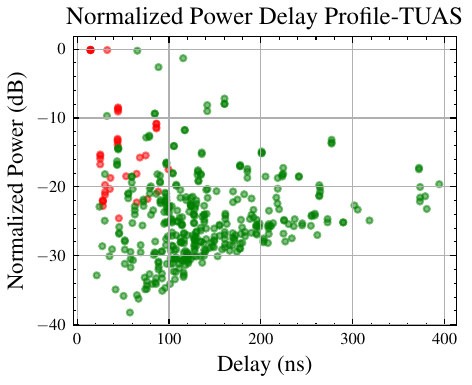}}
    \subfloat[TUAS2]{%
        \includegraphics[scale=0.5]{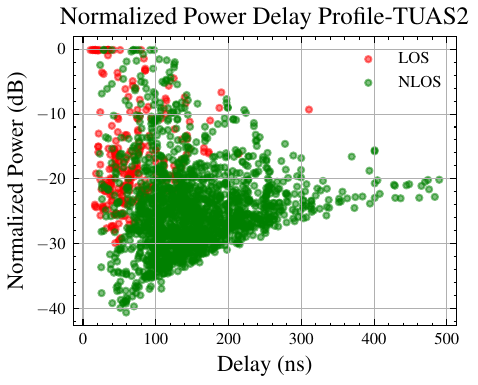}}
    \caption{Normalized power delay profile of indoor locations.}
    \label{fig:PDP} 
\end{figure}

\begin{figure}[t]
    \centering
    \subfloat[Campus]{%
       \includegraphics[scale=0.5]{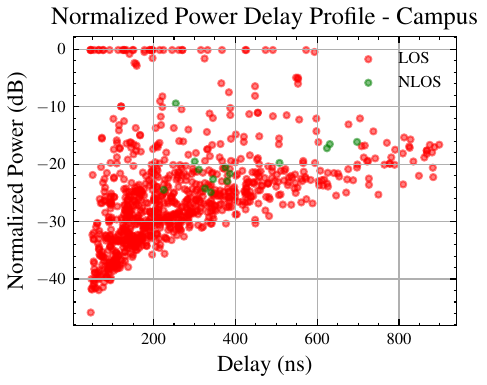}}
    \subfloat[City]{%
        \includegraphics[scale=0.5]{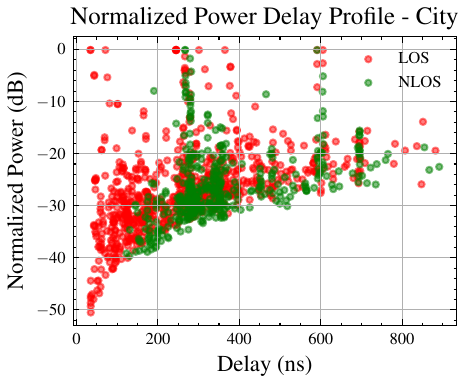}}
    \\
    \subfloat[Residential]{%
    \centering
        \includegraphics[scale=0.5]{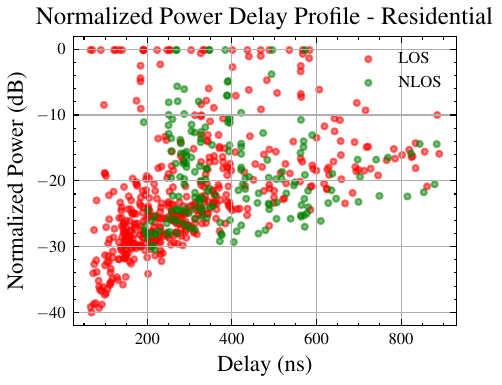}}
    
    \caption{Normalized power delay profile of outdoor locations.}
    \label{fig:PDP-Outdoor} 
\end{figure}

The \ac{PDP} plays a pivotal role in understanding the behaviour of \ac{MPC}s propagation in D band, as it provides a proper understanding on how the power of \ac{MPC} components vary with the time of arrival with phenomena such as diffraction, scattering, and reflection available in the D band. We have analysed the normalized power according to (\ref{equ:power_norm}), against the measured delay to eliminate the impact of \ac{FSPL}. Figs \ref{fig:PDP}  and \ref{fig:PDP-Outdoor} presents the \ac{PDP} of indoor and outdoor locations respectively. 
 
By observing the PDP of indoor and outdoor locations, it is evident that the \ac{PDP}s follow the same pattern for both scenarios.   There are several data points with 0 dB power indicating that the only  loss available in these locations were \ac{FSPL}. In general, these data points should appear in \ac{LOS} measurements. However, interestingly it is visible that there are several \ac{NLOS} data points with 0 dB power after normalization implying that there are \ac{LOS} paths in even in \ac{NLOS} scenarios. The reason for these data points could be  that there are strong reflecting surfaces which will allow the signal to propagate to the receiver with almost zero excess loss in the signal. Further, when we analyze the \ac{PDP} of locations with higher number of data points, such as TUAS, and  TUAS2 for indoor, campus, residential, and city for outdoor we can observe a clear lower bond in the \ac{PDP}. This can be an outcome of the environmental noise dominating over the received power when the transmission distance is high, and the system's inability to capture it due to the limitation of its dynamic range.


\subsection{Normalized Power Distribution (NPD)}

\begin{figure} [t]
    \centering
  \subfloat[Sello]{%
       \includegraphics[scale=0.5]{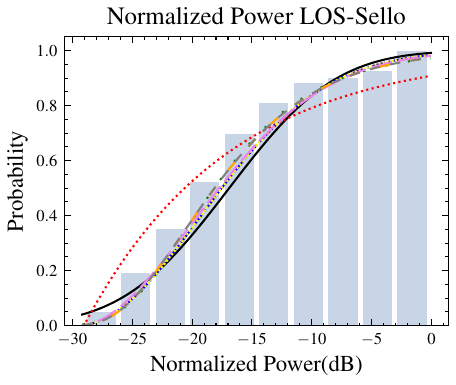}}
    \hfill
  \subfloat[Airport]{%
        \includegraphics[scale=0.5]{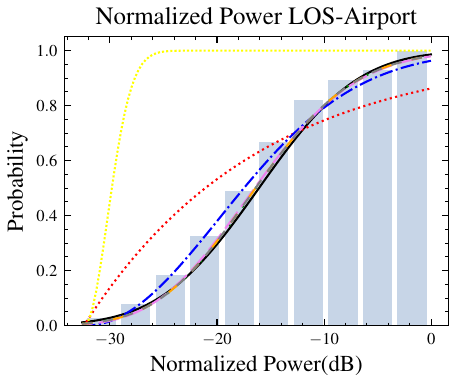}}
    \\
  \subfloat[TUAS]{%
        \includegraphics[scale=0.5]{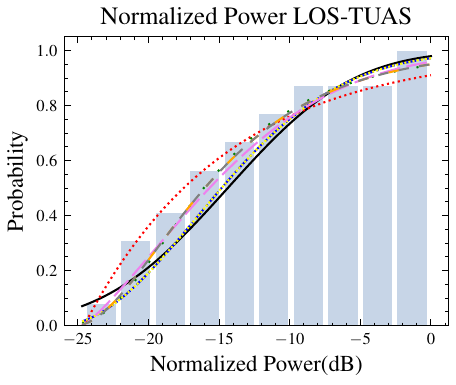}}
    \hfill
  \subfloat[TUAS2]{%
        \includegraphics[scale=0.5]{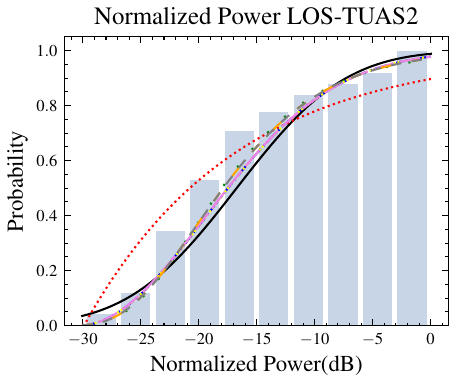}}
         \vspace{0.1mm} 
    \subfloat{%
        \includegraphics[width=0.497\textwidth]{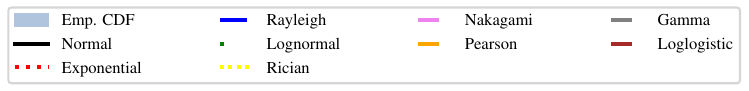}}
  \caption{Normalized power distribution for line-of-sight scenario in indoor locations.}
  \label{fig:NPD - LOS} 
\end{figure}

\begin{figure} [t]
    \centering
  \subfloat[Campus]{%
       \includegraphics[scale=0.5]{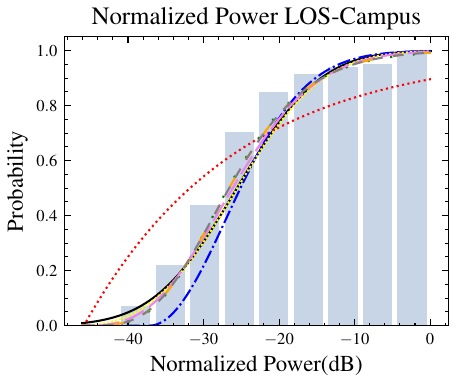}}
    \hfill
  \subfloat[City]{%
        \includegraphics[scale=0.5]{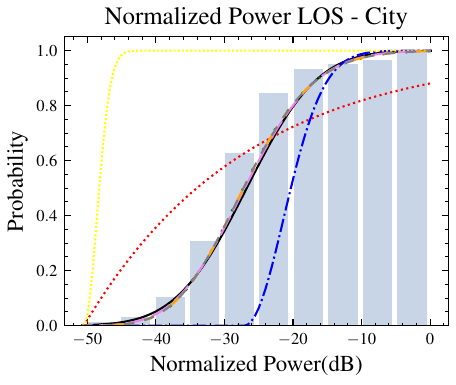}}
    \\
  \subfloat[Residential]{%
  \centering
        \includegraphics[scale=0.5]{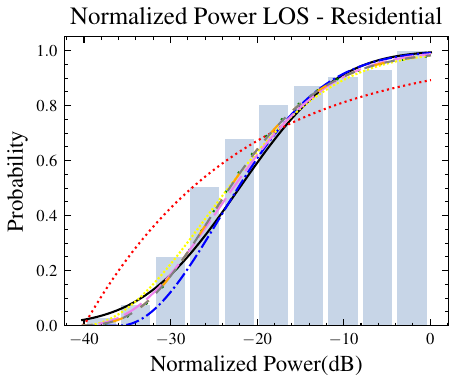}}
    \\
   \vspace{0.1mm} 
    \subfloat{%
        \centering\includegraphics[width=0.497\textwidth]{Results/Number_of_Paths/Indoor/LOS/Fitting/legend.pdf}}
  \caption{Normalized power distribution for line-of-sight scenario in outdoor locations.}
  \label{fig:NPD - LOS-Outdoor} 
\end{figure}

\begin{figure} [t]
    \centering
  \subfloat[Sello]{%
       \includegraphics[scale=0.5]{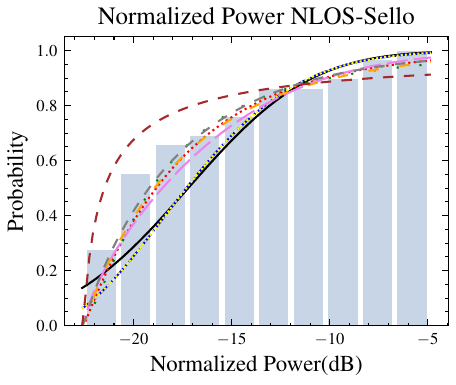}}
    \hfill
  \subfloat[Airport]{%
        \includegraphics[scale=0.5]{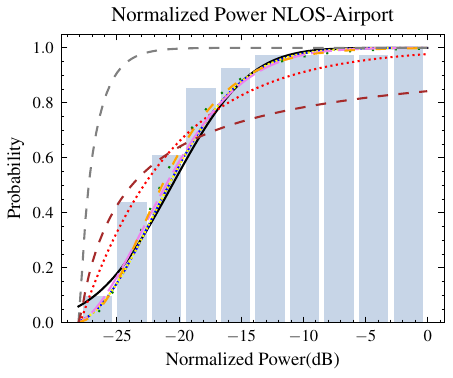}}
    \\
  \subfloat[TUAS]{%
        \includegraphics[scale=0.5]{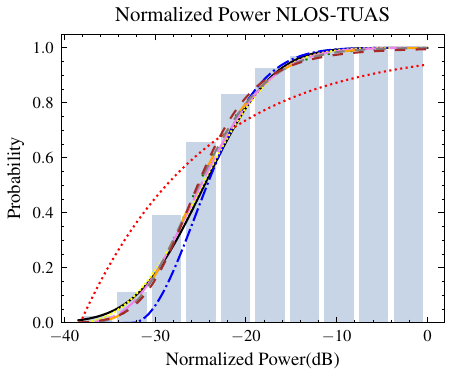}}
    \hfill
  \subfloat[TUAS2]{%
        \includegraphics[scale=0.5]{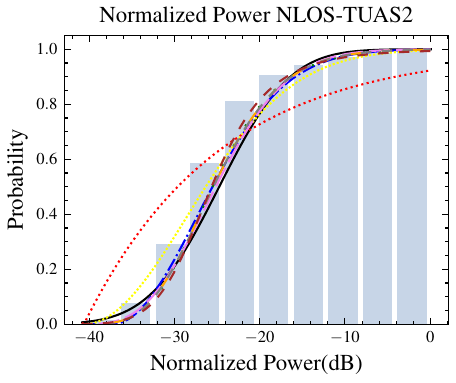}}
         \vspace{0.1mm} 
    \subfloat{%
        \includegraphics[width=0.497\textwidth]{Results/Number_of_Paths/Indoor/LOS/Fitting/legend.pdf}}
  \caption{Normalized power distribution for non-line-of-sight scenario in indoor locations.}
  \label{fig:NPD - NLOS} 
\end{figure}

\begin{figure} [t]
    \centering
  \subfloat[Campus]{%
       \includegraphics[scale=0.5]{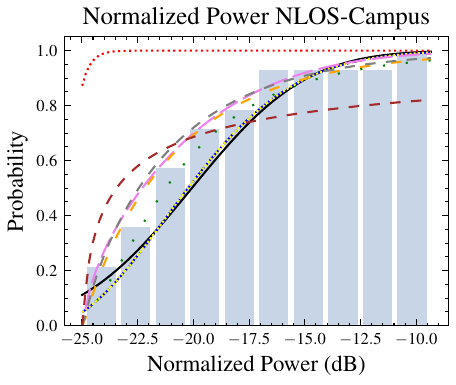}}
    \hfill
  \subfloat[City]{%
        \includegraphics[scale=0.5]{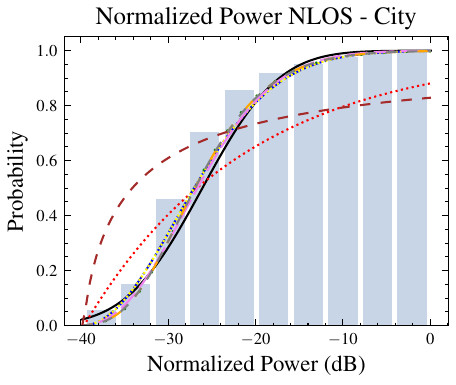}}
    \\
  \subfloat[Residential]{%
  \centering
        \includegraphics[scale=0.5]{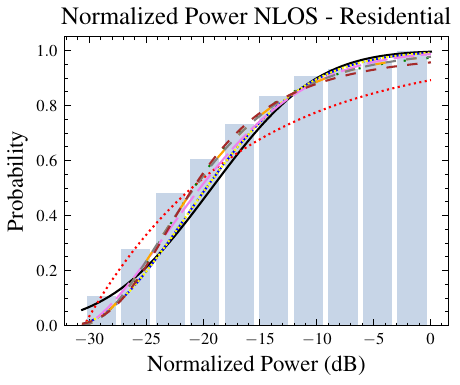}}
    \\
  \vspace{0.1mm} 
    \subfloat{%
        \includegraphics[width=0.497\textwidth]{Results/Number_of_Paths/Indoor/LOS/Fitting/legend.pdf}}
  \caption{Normalized power distribution for non-line-of-sight scenario in outdoor locations.}
  \label{fig:NPD - NLOS-Outdoor} 
\end{figure}

To perceive the distribution of normalized power of \ac{MPC} components, we have analyzed the empirical distribution of the measured data against a set of candidate theoretical distributions to test the compatibility between the two distributions. As mentioned earlier, we have utilized both visual and normality tests to verify the goodness of fit of these distributions. Since the \ac{KS} test compares the \ac{CDF} of the two distributions, we have utilized the \ac{CDF} for visual inspections as well. Furthermore, we have calculated the correlation coefficient with respect to Q-Q plots to understand the compatibility of the two distributions concerning the plots. \Crefrange{fig:NPD - LOS}{fig:NPD - NLOS-Outdoor} present the \ac{NPD} of \ac{LOS} and \ac{NLOS} scenarios for indoor and outdoor measured values, and summary statistics related to distribution fitting and \ac{KS} tests are provided in Appendix \ref{FirstAppendix}.


When we examine both visualizations and goodness of fit statistics in this analysis, it is challenging to effectively fit distributions for locations that have low numbers of data points/paths, such as LOS-TUAS, NLOS-Sello, NLOS-Airport and NLOS- City. The p-values associated with most distributions in these locations either exceed the significance threshold or has very low p-values, indicating insufficient evidence to reject or accept the null hypothesis of goodness of fit. However, visual inspection suggests that many distributions do not fit the data well in these locations. This discrepancy highlights the limitations of relying solely on statistical measures, especially in cases where the dataset size is small.
Nevertheless, considering the entirety of visualizations and goodness of fit statistics reveals that several distributions, including \textbf{log-normal}, \textbf{Nakagami}, \textbf{gamma}, and \textbf{beta} distributions, exhibit satisfactory conformity with the \ac{LOS} scheme for the indoor scenarios while \textbf{log-normal} and \textbf{gamma} distributions had a better fit when outdoor scenarios were considered. Notably, these distributions demonstrate acceptable values for both p and R values, indicating their suitability for modeling \ac{LOS} propagation characteristics when a considerable number of data points are available. Conversely, in the case of the \ac{NLOS} scheme, the data aligns well with the \textbf{log-logistic} distribution for indoor scenarios, particularly when a significant number of data points are available. When outdoor scenarios are considered the \textbf{log-normal} distributions seems to be the best fit for all three locations. 

\subsection{Normalized Delay Distribution }

\begin{figure} [t]
    \centering
  \subfloat[Sello]{%
       \includegraphics[scale=0.5]{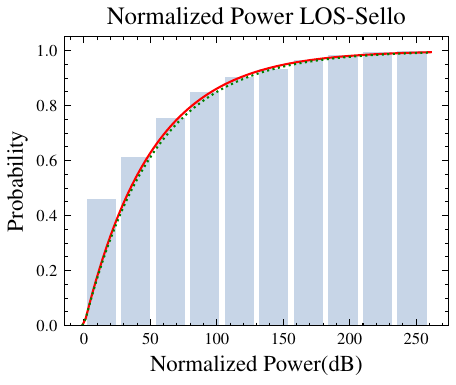}}
    \hfill
  \subfloat[Airport]{%
        \includegraphics[scale=0.5]{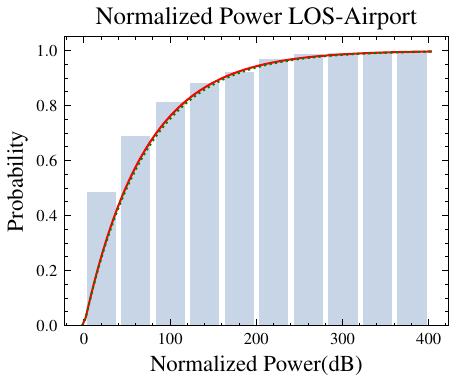}}
    \\
  \subfloat[TUAS]{%
        \includegraphics[scale=0.497]{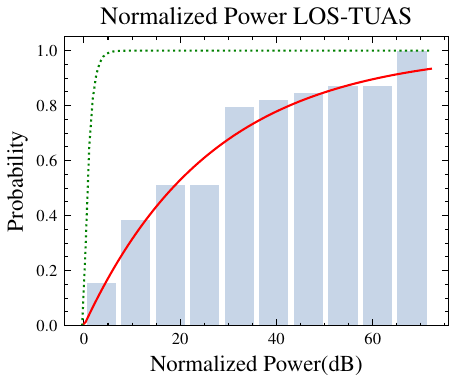}}
    \hfill
  \subfloat[TUAS2]{%
        \includegraphics[scale=0.497]{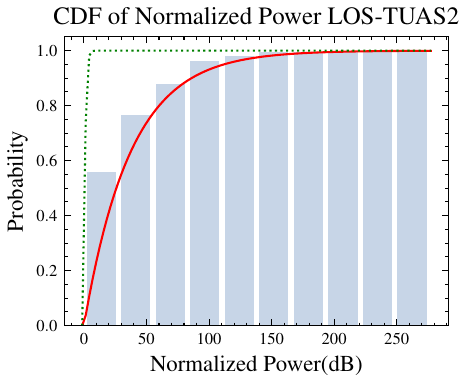}}
        \vspace{0.1mm} 
    \subfloat{%
        \includegraphics[width=0.497\textwidth]{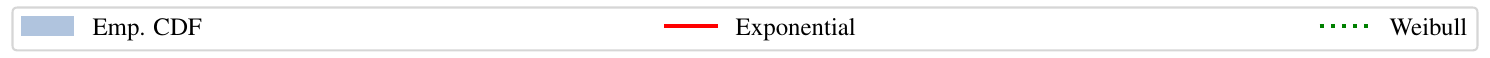}}
  \caption{Normalized delay distribution for line-of-sight scenario in indoor locations.}
  \label{fig:NDD - LOS} 
\end{figure}

\begin{figure} [t]
    \centering
  \subfloat[Campus]{%
       \includegraphics[scale=0.5]{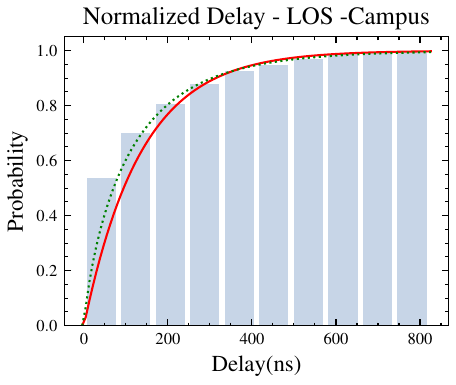}}
    \hfill
  \subfloat[City]{%
        \includegraphics[scale=0.5]{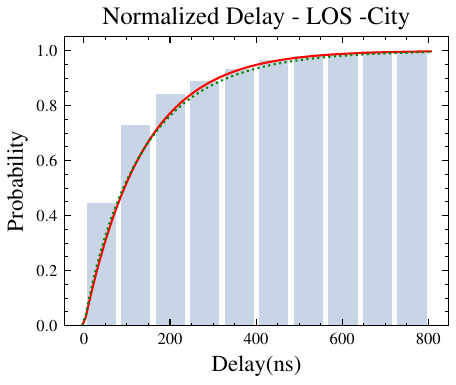}}
    \\
  \subfloat[Residential]{%
  \centering
        \includegraphics[scale=0.5]{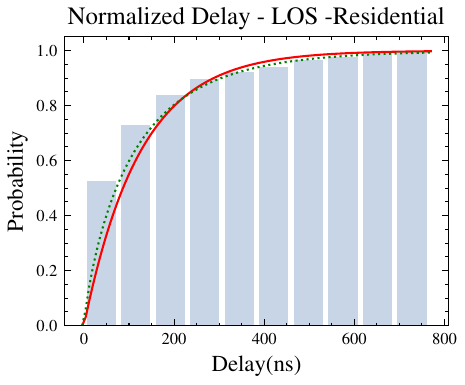}}
    \\
 \vspace{0.1mm} 
    \subfloat{%
        \includegraphics[width=0.497\textwidth]{Results/Delay_Distribution/legend_delay.pdf}}
  \caption{Normalized delay distribution for line-of-sight scenario in outdoor locations.}
  \label{fig:NDD - LOS-outdoor} 
\end{figure}

\begin{figure} [t]
    \centering
  \subfloat[Sello]{%
       \includegraphics[scale=0.5]{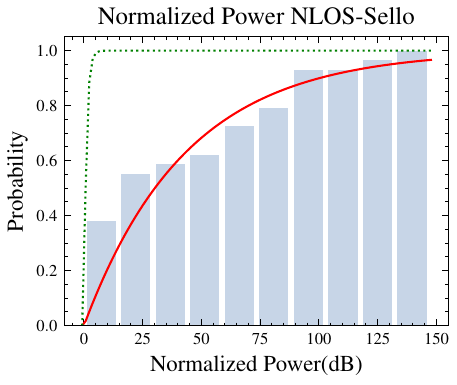}}
    \hfill
  \subfloat[Airport]{%
        \includegraphics[scale=0.5]{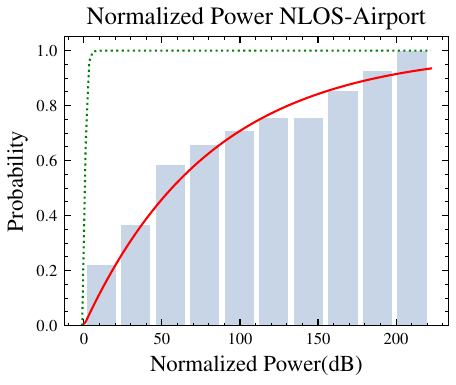}}
    \\
  \subfloat[TUAS]{%
        \includegraphics[scale=0.5]{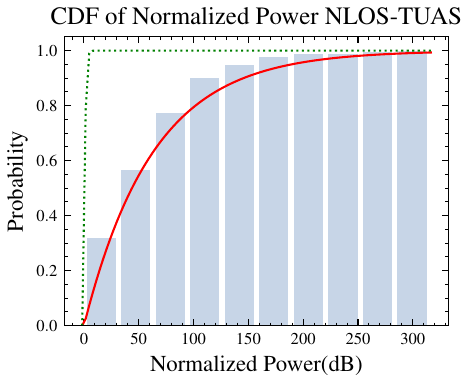 }}
    \hfill
  \subfloat[TUAS2]{%
        \includegraphics[scale=0.5]{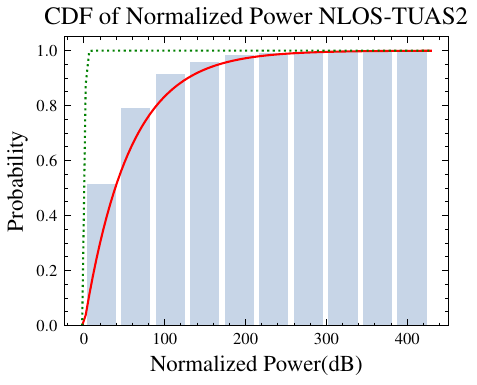}}
        \vspace{0.1mm} 
    \subfloat{%
        \includegraphics[width=0.497\textwidth]{Results/Delay_Distribution/legend_delay.pdf}}
  \caption{Normalized delay distribution for non-line-of-sight scenario in indoor locations.}
  \label{fig:NDD - NLOS} 
\end{figure}

\begin{figure} [t]
    \centering
  \subfloat[Campus]{%
       \includegraphics[scale=0.5]{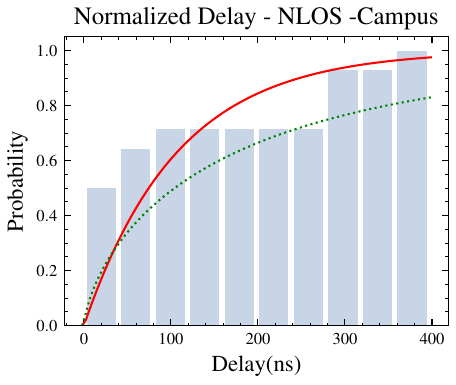}}
    \hfill
  \subfloat[City]{%
        \includegraphics[scale=0.5]{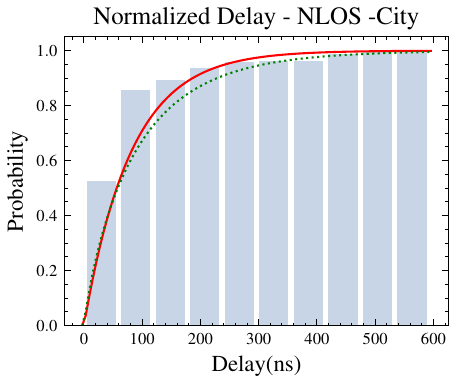}}
    \\
  \subfloat[Residential]{%
  \centering
        \includegraphics[scale=0.5]{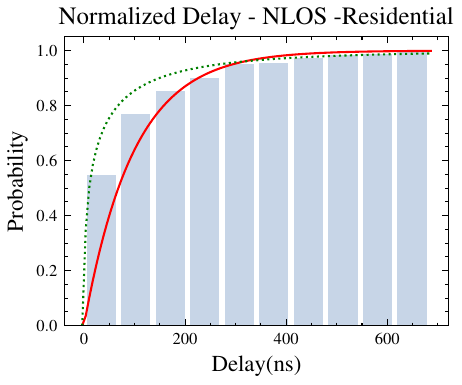}}
    \\
  \vspace{0.1mm} 
    \subfloat{%
        \includegraphics[width=0.497\textwidth]{Results/Delay_Distribution/legend_delay.pdf}}
  \caption{Normalized delay distribution for non-line-of-sight scenario in outdoor locations.}
  \label{fig:NDD - NLOS-Outdoor} 
\end{figure}

To anticipate the delay distribution of the multipath components at D band we have plotted the \ac{CDF}s related to the normalized delay as stated in (\ref{equ: delay_norm})  and compared the results with different theoretical \ac{CDF}s. The resulting graphs are presented in \Crefrange{fig:NDD - LOS} {fig:NDD - NLOS-Outdoor}; the summary statistics related to the distribution fitting and KS test are presented in Appendix \ref{SecondAppendix}.


Similar to the challenges encountered in fitting the distributions to the \ac{NPD}, the \ac{NDD} also presents corresponding difficulties, particularly when the number of data points is low. However, unlike the \ac{NPD}, the \ac{NDD} reveals a limited number of distributions suitable for empirical fitting. Specifically, the Weibull and exponential distributions emerge as potential candidates for modeling the \ac{NDD} behavior. Upon comprehensive analysis of statistics and visualizations, it becomes evident that the \textbf{exponential} distribution emerges as the most suitable fit for both \ac{LOS} and \ac{NLOS} schemes for both indoor and outdoor locations within the \ac{NDD} framework. It offers the most accurate representation of delay characteristics in both \ac{LOS} and \ac{NLOS} scenarios despite the challenges posed by limited dataset sizes.

\subsection{The Number of Paths }

\begin{figure} [t]
    \centering
  \subfloat[Sello]{%
       \includegraphics[scale=0.497]{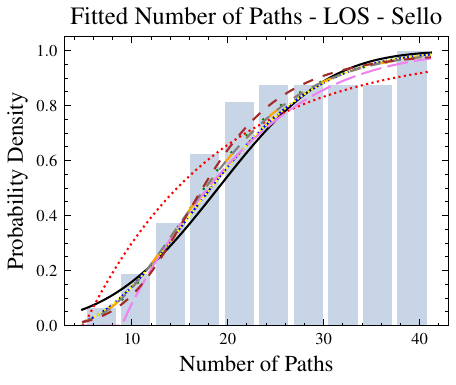}}
    \hfill
  \subfloat[Airport]{%
        \includegraphics[scale=0.497]{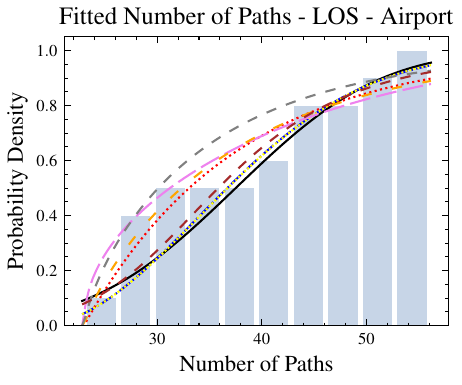}}
    \vspace{0.1mm} 
    \subfloat{%
        \includegraphics[width=0.497\textwidth]{Results/Number_of_Paths/Indoor/LOS/Fitting/legend.pdf}}
  \caption{Distribution of \ac{MPC}s for line-of-sight scenario in indoor locations.}
  \label{fig : NoP-LOS-fitting}
\end{figure}

\begin{figure} [t]
    \centering
  \subfloat[TUAS]{%
        \includegraphics[scale=0.497]{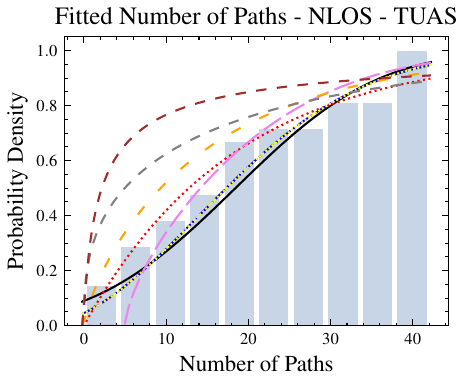}}
    \hfill
  \subfloat[TUAS2]{%
        \includegraphics[scale=0.497]{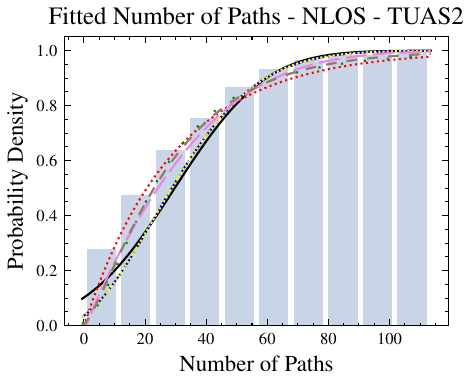}}
        \vspace{0.1mm} 
    
    \subfloat{%
        \includegraphics[width=0.497\textwidth]{Results/Number_of_Paths/Indoor/LOS/Fitting/legend.pdf}}

  \caption{Distribution of \ac{MPC}s for non-line-of-sight scenario in indoor locations.}
  \label{fig : NoP-NLOS-fitting} 
\end{figure}

\begin{figure} [t]
    \centering
  \subfloat[Campus]{%
       \includegraphics[scale=0.5]{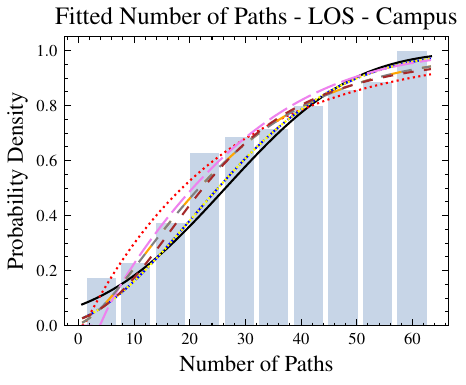}}
    \hfill
  \subfloat[Residential]{%
        \includegraphics[scale=0.5]{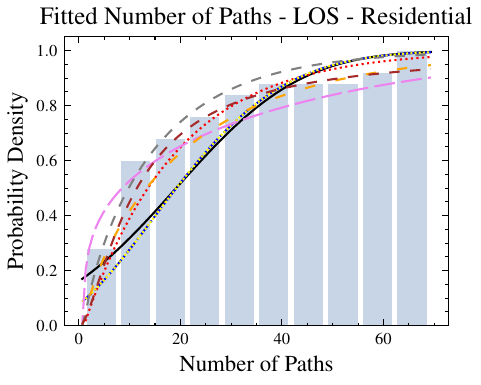}}
         \vspace{0.1mm} 
    
    \subfloat{%
        \includegraphics[width=0.497\textwidth]{Results/Number_of_Paths/Indoor/LOS/Fitting/legend.pdf}}
  \caption{Distribution of \ac{MPC}s for line-of-sight scenario in outdoor locations.}
  \label{fig : NoP-LOS-fitting-outdoor}
 
\end{figure}

\begin{figure} [t]
    \centering
  \subfloat[City]{%
        \includegraphics[scale=0.5]{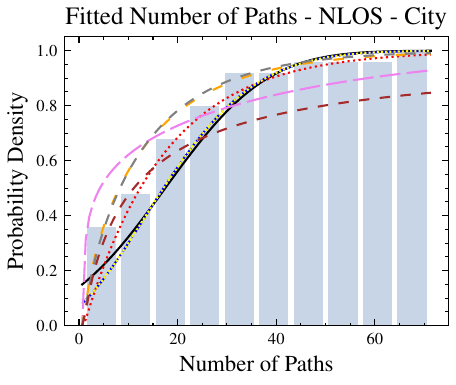}}
    \hfill
  \subfloat[Residential]{%
        \includegraphics[scale=0.5]{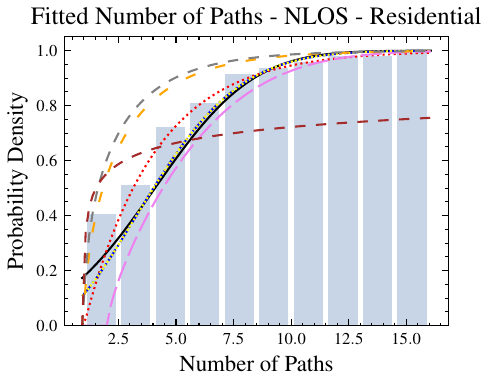}}
        \vspace{0.1mm} 
    
    \subfloat{%
        \includegraphics[width=0.497\textwidth]{Results/Number_of_Paths/Indoor/LOS/Fitting/legend.pdf}}
  \caption{Distribution of \ac{MPC}s for non-line-of-sight scenario in outdoor locations.}
  \label{fig : NoP-NLOS-fitting-outdoor} 
  
\end{figure}

\begin{figure} [t]
    \centering
  \subfloat[Sello]{%
       \includegraphics[scale=0.45]{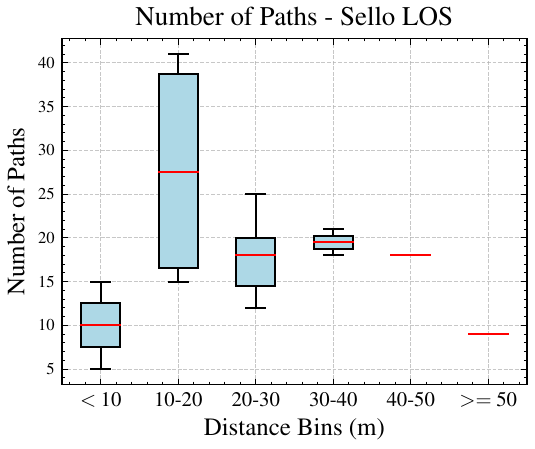}}
    \hfill
  \subfloat[Airport]{%
        \includegraphics[scale=0.45]{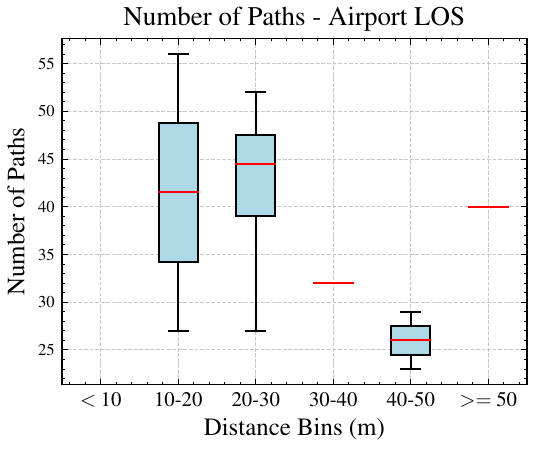}}
  \caption{\ac{MPC}s vs. Tx-Rx  distance for line-of-sight in indoor locations}
  \label{fig : NoP-LOS}
\end{figure}

\begin{figure} [t]
    \centering
  \subfloat[TUAS]{%
        \includegraphics[scale=0.45]{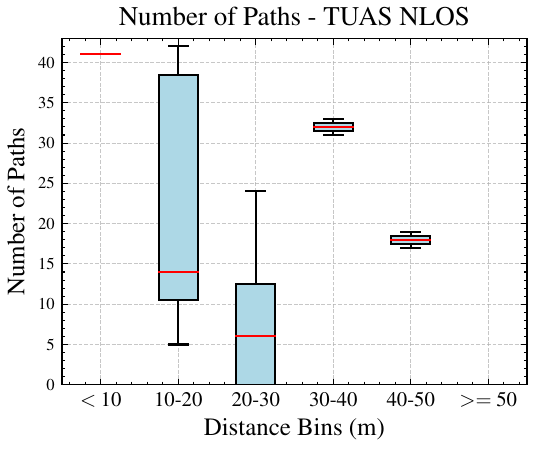}}
    \hfill
  \subfloat[TUAS2]{%
        \includegraphics[scale=0.45]{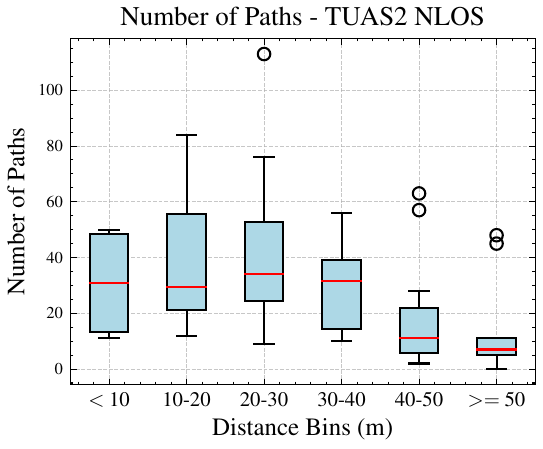}}
  \caption{\ac{MPC}s vs. Tx-Rx  distance for non-line-of-sight in indoor locations}
  \label{fig : NoP-NLOS} 
\end{figure}

\begin{figure} [t]
    \centering
  \subfloat[Campus]{%
       \includegraphics[scale=0.45]{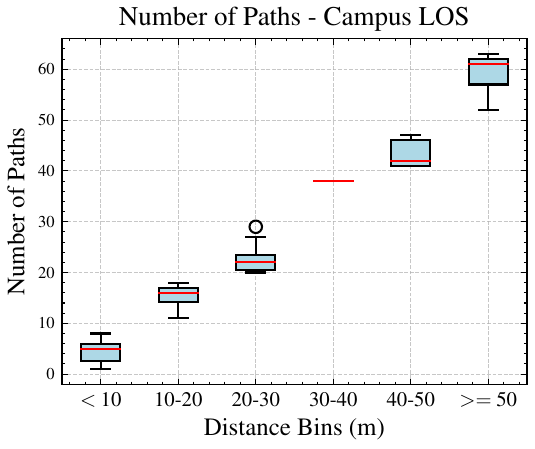}}
    \hfill
  \subfloat[Residential]{%
        \includegraphics[scale=0.45]{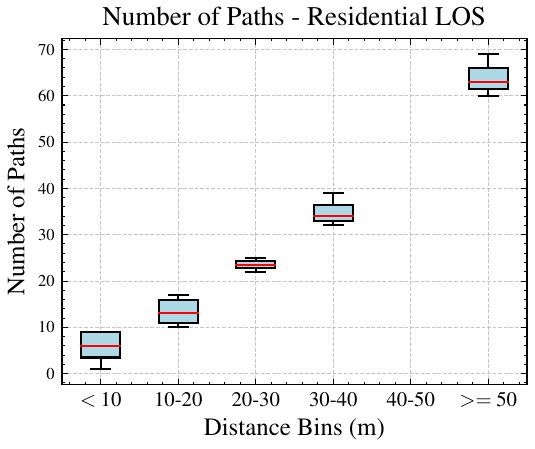}}
  \caption{\ac{MPC}s vs. Tx-Rx  distance for line-of-sight in outdoor locations.}
  \label{fig : NoP-LOS-outdoor}
\end{figure}

\begin{figure} [t]
    \centering
  \subfloat[City]{%
        \includegraphics[scale=0.45]{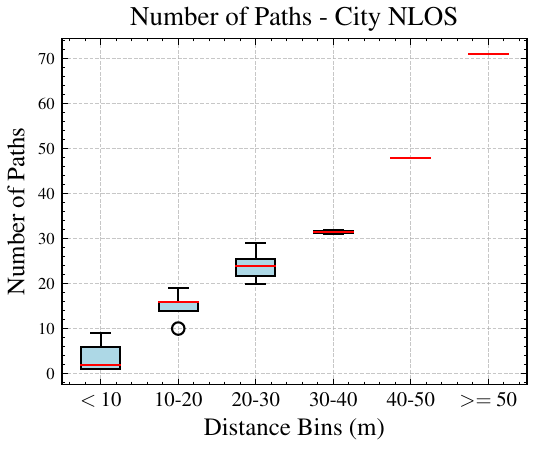}}
    \hfill
  \subfloat[Residential]{%
        \includegraphics[scale=0.45]{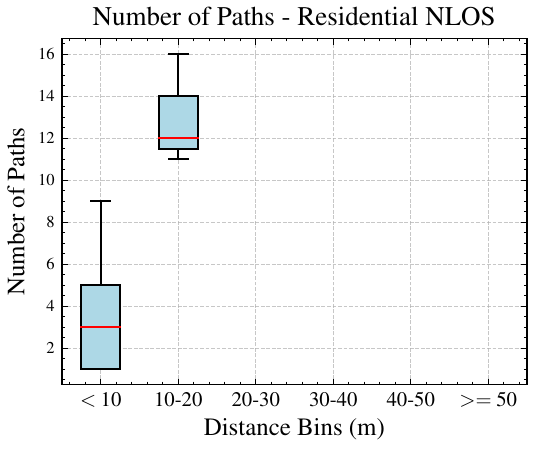}}
  \caption{\ac{MPC}s vs. Tx-Rx  distance for non-line-of-sight in outdoor locations.}
  \label{fig : NoP-NLOS-Outdoor} 
\end{figure}

To discern the \ac{NoP} in D band indoor and outdoor wireless channels, we have analysed the data in two perspectives. Firstly, we have analyzed the \ac{NoP} to examine which distribution best fits the empirical data, similar to the analysis in \cite{hashemi1993impulse}. Secondly, we have analysed the \ac{NoP} captured for each distance measurement between the \ac{Tx} and the \ac{Rx}. The visualizations related to this analysis are presented in \Crefrange{fig : NoP-LOS-fitting}{fig : NoP-NLOS-Outdoor}. The maximum, minimum, and mean  of \ac{NoP}s for each location are provided in Table \ref{tab:Stats-NoP}, and summary statistics of the KS test for different distributions applied to this data set are provided in Appendix \ref{ThirdAppendix}. For the \ac{LOS} scenario, we considered the Sello, Airport indoor locations  and Campus, Residential outdoor locations. TUAS, TUAS2 and City, Residential locations were selected for the \ac{NLOS} scenario for indoor and outdoor locations respectively. This location selection was based on the highest number of measurement points available as mentioned in Table \ref{tab:Number of Data Points}.





\begin{table}[t]
\caption{Statistics of Number of Paths}

    \label{tab:Stats-NoP}
    \centering
\begin{tabular}{l| c c c|c c c} \hline
     \multirow{2}{*}{Location } &\multicolumn{3}{c}{LOS} & 
        \multicolumn{3}{c}{NLOS} \\ 
         & Max. & Min. & Mean & Max. & Min. & Mean \\ \hline
        Sello&41 & 5 & 19 &19 &10 & 14.5 \\ 
        Airport & 56 &23 & 37.5 & 41& 41 &41\\ 
         TUAS & 11 &6& 7.8& 42& 0 & 18.5\\ 
        TUAS2 & 49 &6 & 16.8& 113 &0&29.7\\ \hline
     Campus&63 & 1 & 26.4 & 4&1&2.25\\ 
 City & 178&8 & 64 & 71&1& 17.48\\
    Residential & 69 &1 & 19.16& 16&1 & 4.01\\ \hline
\end{tabular}
\end{table}
By carefully examining the statistics of the of goodness of fit, it is evident that log-logistic distribution best describes the distribution of the \ac{NoP} when the \ac{LOS} scenario is considered for both indoor and outdoor locations and normal distribution fits best when \ac{NLOS} data are considered. However, when we cross examine the visualisations and the goodness of fit statistics is becomes evident that we cannot rely on these conclusion as they contradict with one another. The main cause for this can be due to the lack of data points for this specific analysis as we are considering the number of measurement points as stated in Table \ref{tab:Number of Data Points}. Furthermore,  the
box plot visualizations of indoor locations (\Crefrange{fig : NoP-LOS}{fig : NoP-NLOS}) suggest
that the median of \ac{NoP} in indoor locations increases up to
the \qty{20}{m} -- \qty{30}{m} distance range, following which it begins to decline
with the \ac{Tx}-\ac{Rx} separation. However, considering the indoor
dataset, a consistent trend is evident. The median \ac{NoP} peaks
when the \ac{Tx}-\ac{Rx} distance falls within the \qty{10}{m} -- \qty{30}{m} distance range
for both \ac{LOS} and \ac{NLOS} scenarios. This phenomenon can be attributed to the positioning of the \ac{Tx} and \ac{Rx} creating very few short paths initially, while the \ac{NoP} increases with increasing distance. Subsequently, when the \ac{Tx} and \ac{Rx} are further separated, the power of these components falls below a specific threshold due to the dynamic range of the system, becoming undetectable to the \ac{Rx}. Unlike indoor locations, the median of \ac{NoP} for outdoor locations increase with \ac{Tx}-\ac{Rx} separation. This trend may be due to the tunneling effect created by the nature of the outdoor environment.

\subsection{Angular Spread}

A detailed analysis on angular spread for the same data set is presented in \cite{de2024comparison}. This analysis reveals that the angular spread only slightly decreases with the frequency band. Further, it is revealed that the angular spread follows a log-normal distribution and the statistics related to each location with a detailed analysis are provided in \cite{de2024comparison}. For the \ac{GBSM} model provided in this study, we assume that the angular spread follows a log-normal distribution. 

The statistical analysis presented in Section \ref{sec:results} provides valuable insights into the characteristics of \ac{mmWave} propagation in the D band. To further validate these findings and assess the practical applicability of our proposed \ac{GBSM}, the next section concentrates on model verification and a comprehensive discussion of the results.

\section{Verifications and Discussion}\label{sec:verfi}

In this section, we will first evaluate the performance of our \ac{GBSM} by comparing its predictions against the empirical data, focusing particularly on the \ac{MED} as a key performance metric. Following this verification, we will engage in a detailed discussion of our findings, exploring their implications for D Band communications.
\subsection{Maximum Excess Delay}

\begin{figure} [t]
    \centering
  
  \subfloat[TUAS2]{%
        \includegraphics[scale=0.5]{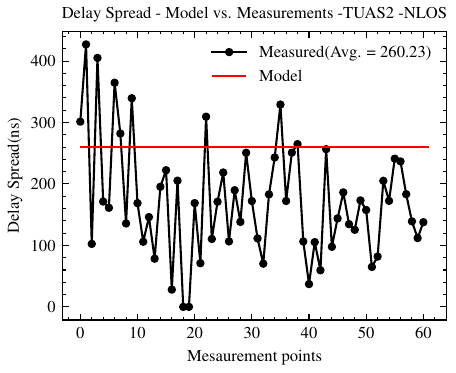}
        }
        \subfloat[Campus]{%
       \includegraphics[scale=0.5]{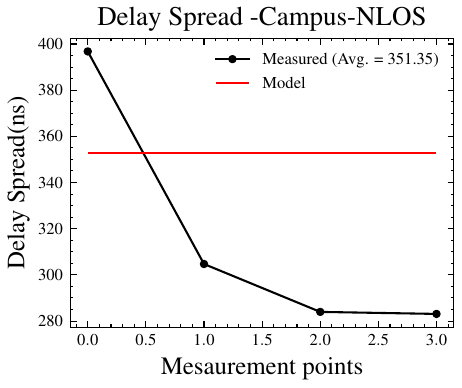}}
  \caption{Delay distribution comparison between the measured data and model predictions. }
  \label{fig:DD} 
\end{figure}

As mentioned in Section \ref{sec:methodology}, we have used \ac{MED} as a metric to evaluate the performance of our model against the empirical data. The detailed description of \ac{MED} calculation based on this model is provided in Section \ref{sub:delay_spread}. Although several candidate distributions were identified for the LOS scenario of NPD in both indoor and outdoor locations, we selected the log-normal distribution for both scenarios in this analysis, based on our consideration of both goodness-of-fit statistics and visualizations. The average \ac{MED} of each location based on empirical data and the model is provided in Table \ref{tab:delay spread}. However, it's worth noting that we have not used any distribution to represent the number of paths, as the results of the analysis are inconclusive. Instead, the \ac{NoP} value related to each measurement from the empirical data has been used. Further to the discovery of the accuracy of the model when compared to the empirical data, this comparison also highlights the impact of the number of data points through out the analysis. As Figure \ref{fig:DD} represents the two scenarios of the delay distribution of empirical and model data in reference to a scenario with higher number of data points and very low number of data points.

\begin{table}[t]
    \centering
    \caption{Comparison of empirical data and model predictions for average \ac{MED} }  \label{tab:delay spread}
    \begin{tabular}{c|c|c|c}\hline
         Location &Scenario & \makecell{Empirical \ac{MED} (ns)} & \makecell{Model \ac{MED} (ns)}  \\ \hline
         \multirow{2}{*}{Sello} & {LOS}& 155.2 & 113.2\\ 
         & {NLOS}& - &  -\\\hline
          \multirow{2}{*}{Airport} & {LOS}& 276.56 & 240.78 \\ 
         & {NLOS}& - &  -\\\hline
           \multirow{2}{*}{TUAS} & {LOS}& 110.69 & 88.13 \\ 
         & {NLOS}& 159.78 &  228.89\\\hline
           \multirow{2}{*}{TUAS2} & {LOS}& 110.69 &  124.54\\ 
         & {NLOS}& 172.65 &  260.31\\\hline
             \multirow{2}{*}{Campus} & {LOS}& 542.25 &  562.04\\ 
         & {NLOS}& 317.12 &  351.35\\\hline
             \multirow{2}{*}{City} & {LOS}& 482.51 & 408.04 \\ 
         & {NLOS}& 281.1 & 306.65 \\\hline
             \multirow{2}{*}{Residential} & {LOS}& 376.49 & 467.87 \\ 
         & {NLOS}& 188.7 &  412.17\\\hline
         
    \end{tabular}

\end{table}


\subsection{Channel Modeling Based on Measurements }

This study has focused on deriving channel statistics purely based on measurement data. There were both benefits and drawbacks to this method. The main benefit is the ability to model the real environment based on measurements rather than relying on assumptions about the environment. This enables more accurate channel characterization, leading to more precise channel modeling.

However, throughout this analysis, it was evident that there were a few scenarios with insufficient data points, leading to inconclusive results (e.g., \ac{NoP}) and further we were unable to analyse specific connection between these parameters due to the same limitation. This is the main drawback of building channel models solely based on measurement data, as the measurement campaign can be time-consuming and may result in an inadequate number of data points for proper analysis.

\subsection{Discussion}

Upon reviewing the results, it becomes apparent that several distributions can be used to represent the \ac{NPD} of the \ac{LOS}-indoor scenario, while the log-normal and gamma distributions seems to be the best fit for the \ac{LOS}-outdoor scenario. When the \ac{NLOS} scenario is considered, the log-logistic distribution appears to be the only eligible candidate to represent the distribution of the \ac{NPD} for indoor locations, while the log-normal distribution has the best statistics for outdoor locations. Unlike the \ac{NPD}, in the \ac{NDD} distribution analysis there are only a few candidate distributions that can be considered to represent the \ac{NDD} distribution, namely the Weibull and exponential distributions. Among these two, the exponential distribution outperforms the Weibull distribution in all scenarios. Furthermore, discrepancies between the results of the \ac{KS} test and Q-Q plots are noted. These inconsistencies may stem from the relatively low number of data points used in the analysis. The \ac{KS} test's sensitivity to dataset size can lead to unreliable performance when data points are scarce, while Q-Q plots may tend to overfit small datasets.

When the number of \ac{MPC} components per transmission is considered, it is difficult to arrive at a conclusion on a specific distribution for its distribution relying solely on the visual representations and the statistics of goodness of fit as they contradict with one another due to the lack of data points. However, when the number of \ac{MPC}s are plotted against the \ac{Tx}-\ac{Rx} separation, a clear pattern can be observed in both indoor and outdoor locations for \ac{LOS} and \ac{NLOS} scenarios.

Considering indoor and outdoor locations, it can be observed that both scenarios have similar patterns in \ac{LOS} transmission, with slight changes in the \ac{NLOS} transmissions on \ac{NPD} distributions. However, when the \ac{NoP} per transmission is considered, it can be observed that the \ac{NoP}s in outdoor locations is reasonably higher compared to indoor transmissions for \ac{LOS} scenario, while the \ac{NoP}s are lower in outdoor locations for the \ac{NLOS} scenario. There can be two valid explanations for this. One is that the measurements of the \ac{MPC} are captured below a 30 dB threshold from the strongest measured path. Since the \ac{Tx}/\ac{Rx} separation of the \ac{LOS} outdoor scenarios is greater than that of the indoor locations, the strongest path of outdoor locations may have less power, resulting in a threshold that might capture more \ac{MPC}s outdoors. Secondly,the indoor environments have more lossy interactions than outdoor environments, reducing the number of captured \ac{MPC}s.


\section{Conclusions }\label{sec:conclusion} 

In this paper, we derived a \ac{GBSM} for \ac{mmWave} communications above 100 GHz relying on  spatial-temporal statistics based on measurements conducted in the D band with the motivation of creating a generalized channel frequency response for \ac{MIMO} communications in \ac{mmWave} frequency. Specifically, a detailed analysis was conducted on path gain, delay, and the \ac{NoP}, considering both \ac{LOS} and \ac{NLOS} scenarios for indoor and outdoor locations. 
Theoretical distributions were utilized to characterize the behavior of \ac{MPC}s for path gain and delay in the D band. While several distributions accurately characterize the behavior of the \ac{LOS} power distribution for indoor locations, other distributions, namely the \ac{NLOS} power distribution for indoor locations characterized by  log-logistic distribution,  \ac{LOS} and \ac{NLOS} \ac{NPD} for outdoor locations through log-normal as well as the \ac{LOS} and \ac{NLOS} delay distributions for all locations characterized by the exponential distribution, exhibit specific characteristics that can only be accurately captured by these relevant distributions.  This comprehensive analysis, together with the analysis on angular spread has enabled us to arrive at a more accurate channel frequency response modeling for the \ac{mmWave} frequency communication.  To confirm the accuracy of our model, we have compared the \ac{MED} of the synthetic data generated from the model against the empirical data which we gathered through measurements. Interestingly, we can observe that the model presented in this research provides acceptable results against the empirical data. However, this research also emphasis the shortcoming of lack of  measurement points for a detailed analysis motivating future research to be conducted on increased number of measurements for higher frequency communications.


\appendices
\section{Normalized power distribution summary statistics} \label{FirstAppendix}

Table \ref{tab:NPD -Indoor- Results - Summary} and \ref{tab:NPD - outdoor- Results - Summary} provides the statistics related to the \ac{KS} test -- an estimate of the maximum distance between the two \ac{CDF}s under testing, the p-value which provides a threshold for acceptance of the theoretical distributions, and the parameters: \textbf{loc} -- shift of the distribution along the x-axis, \textbf{scale} -- the standard deviation of the distribution, and the \textbf{shape} parameter which controls the skewness of the distribution while fitting the measured data to each candidate distribution for indoor and outdoor locations respectively. 
\begin{table*} [t]
    \centering
    \caption{Summary statistics of the normalized power distribution in indoor locations - (compatible with python scipy.stats library)}
    \label{tab:NPD -Indoor- Results - Summary}
    
    \begin{tabular}{l |l| cccccc| cccccc}
    \hline
    \multirow{2}{*}{Location} &
    \multirow{2}{*}{Distribution} &
        \multicolumn{5}{c}{LOS} & 
        \multicolumn{5}{c}{NLOS} \\ 
    & & KS Stat &  p-value & R & Loc &  Scale & Shape & KS Stat & p-value & R &  Loc &  Scale & Shape\\ \hline
    
   \multirow{9}{*}{Sello} & Normal      & 0.081 & 0.033 & 0.97 & -17  & 7 & -  &0.223 &0.096  & 0.92 & -17.2 & 4.93  & - \\
    & Exponential & 0.205 & \num{9e-12} & 0.97 &  -29 &  12.1 & - &0.106 & 0.866 & 0.96 & -22.5  & 5.31  & -  \\
    & Log-Normal & \textbf{0.344} & \textbf{0.853} & \textbf{0.99} & -35.5  & 17.4  & 0.37&0.120   & 0.751 & 0.91 & -23.1  & 4.03  & 0.91  \\
    & Rayleigh  & 0.048 & 0.446 & 0.99 & -29.6  & 10.3  & - &0.232& 0.074 & 0.96 & -24.9  & 6.48  & -  \\
    & Rician & 0.048 & 0.446 & 0.99 & -29.6  & 10.3 & 0&0.232  & 0.074 & 0.96 & -24.9 & 6.48 & 0 \\
    & Nakagami & 0.046 & 0.524 & 0.99 & -29.3   & 14.2  & 0.876 &0.118  & 0.772  & 0.98 & -22.5  & 7.26 & 0.88  \\
    & Gamma  & \textbf{0.036} & \textbf{0.811} & \textbf{0.99}&-30.7  & 3.73  & 3.66&0.103  & 0.721 & 0.98 & -22.5  & 5.74  & 0.88  \\
    & Beta &  \textbf{0.036} &\textbf{0.811} & \textbf{0.99} & -16.9& 7.14 & 1.04 &0.127  & 0.883 & 0.98 & -17.1   & 5.6  & 2.1  \\
    & Log Logistic & 0.35 & \num{2e-34} & 0.57 &  -29   & 5.6  & 0.84  &0.413 & \num{5e-05} & 0.58 & -22.5 & 1.1  & 0.83  \\ \hline
    
    \multirow{9}{*}{Airport} & Normal  &0.036     & 0.689 & 0.99 & -16.08  & 7.3 & - &0.124  & 0.5092& 0.92 & -20.5 & 4.83  & - \\
    & Exponential &0.26 & \num{2e-24} & 0.93 &  -32.4 &  16.3 & - &0.241  & 0.014 & 0.97 & -27.9  & 7.38  & -  \\
    & Log-Normal &\textbf{0.029}  & \textbf{0.884} & \textbf{0.99} & -105.5 & 89.16  & 0.08 & 0.087  & 0.892 & 0.97 & -30.9  & 9.4 & 0.42  \\
    & Rayleigh &0.087  & 0.005 & 0.99 & -32.2  & 16.3  & - &0.105& 0.715 & 0.96 & -28.75  & 6.72  & -  \\
    & Rician &0.908 & 0 & 0.27  & -32.4  & 2.2 & 0 &0.105 & 0.715 & 0.95 & -28.7  & 6.7 & 0  \\
    & Nakagami &\textbf{0.031}  & \textbf{0.839} & \textbf{0.99} & -41.2   & 26.1  & 3.04 &0.09 & 0.864 & 0.96 & -28.2  & 9.06 & 0.79  \\
    & Gamma &\textbf{0.029}  & \textbf{0.892} & \textbf{0.99} &-72.5  & 0.94 & 59.7 &0.072 & \num{1e-31} & 0.97 & -27.9 & 1.55  & 0.73  \\
    & Beta &0.029 & 0.892 & 0.99 & -16 & 7.31 & 0.26  &0.827& 0.971 & 0.97 & -20.5   & 4.5  & 1.1  \\
    & Log Logistic  &0.035 & 0.72 & 0.99 &  -102.7  & 86.3  & 20.6 &0.334 & \num{7e-05} & 0.78 & -27.9 & 4.14 & 0.88  \\ \hline
    
    \multirow{9}{*}{TUAS} & Normal   &0.115   & 0.635 & 0.95 & -14.3  & 7 & - &0.074 & 0.026 & 0.97 & -24.5 & 5.8 & - \\
    & Exponential&0.152 & 0.297 & 0.93&  -24.5 &  10.1 & - &0.320 & \num{7e-36} & 0.90 & -38.2  & 13.67  & -  \\
    & Log-Normal  &\textbf{0.120}& \textbf{0.965} & \textbf{0.94} & -28.2  & 12.1  & 0.52 &0.027 & 0.923 & 0.96 & -48.56  & 23.3  & 0.24  \\
    & Rayleigh &0.107  & 0.725 & 0.97  & -26.4  & 9.8  & - &0.119& \num{3e-05} & 0.96 & -32.5  & 6.9 & -  \\
    & Rician&0.107& 0.725 & 0.97 & -26.4  & 9.8 & 0 &0.06 & 0.112 & 0.96 & -38.5  & 6.6 & 1.8  \\
    & Nakagami &\textbf{0.086} & \textbf{0.907} & \textbf{0.97} & -24.7   & 12.4 & 2.1&0.048 & 0.381 & 0.96 & -39.5  & 16.0  & 1.78  \\
    & Gamma  &\textbf{0.069}& \textbf{0.958} & \textbf{0.96} &-25.2  & 5.04  & 1.36 &0.031 & 0.831 & 0.99 & -42.47  & 1.88  & 9.50  \\
    & Beta &\textbf{0.069}& \textbf{0.958} & 0.96 & -14.4 & 7.4& 1.04 &0.031 & 0.831 & 0.99 & -24.5  & 5.8  & 0.64  \\
    & Log Logistic &0.04 & \num{2.3e-05} & 0.47 &  -24.6   & 3.3  & 0.81&\textbf{0.01}  & \textbf{0.969}
 & \textbf{0.99} & -46.11 & 20.78  & 6.62  \\ \hline
 
    \multirow{9}{*}{TUAS2} & Normal  & 0.126    & \num{3e-04} & 0.94 & -16.7  & 7.3 & -  & 0.084& \num{9e-12}
 & 0.98 & -24.63 & 6.5  & - \\
    & Exponential & 0.246& \num{7e-15} & 0.90 &  -29.9 &  13.1 & - & 0.33 & \num{8 e-177} & 0.24 & -40.6  & 15.8  & -  \\
    & Log-Normal & \textbf{0.059} & \textbf{0.283} & \textbf{0.98} & -36.4  & 18.3  & 0.37 & 0.042  & \num{2e-03} & 0.98 & -50.1  & 24.6  & 0.24  \\
    & Rayleigh & 0.093  & 0.017 & 0.98 & -30.42  & 11  & - & 0.059& \num{5e-06} & 0.98 & -37.3 & 10  & -  \\
    & Rician & 0.093 & 0.017 & 0.98 & -30.42  & 11  & 0 & 0.142& \num{3e-32} & 0.98 & -40.6 & 12.1 & 0.11  \\
    & Nakagami  & \textbf{0.093}& \textbf{0.018} & \textbf{0.98} & -30.3  & 15.5 &  1 & 0.064 & \num{5e-07} & 0.97 & -41.7  & 18.1  & 1.8 \\
    & Gamma  & \textbf{0.069}& \textbf{0.146} & \textbf{0.99} &-32.0  & 3.51  & 4.3 & 0.048 & \num{3e-04} & 0.98 & -44.4  & 2.1  & 9.56 \\
    & Beta & 0.069& \num{4e-04} & 0.93 & -16.7 & 7.3 & 0.95  &0.048 & \num{4e-04} & 0.98 & -24.7 & 6.3 & 0.64  \\
    & Log Logistic& 0.042  & 0.70 & 0.99 &  -33.7   & 15.5  & 4.04  &\textbf{0.017}& \textbf{0.878} & \textbf{0.98} & -47.4 & 21.7  & 6.4  \\ \hline
    \end{tabular}
    
\end{table*}

\begin{table*} [t]
    \centering
    \caption{Summary statistics of the normalized power distribution in outdoor locations- (compatible with python scipy.stats library)}
    \label{tab:NPD - outdoor- Results - Summary}
    
    \begin{tabular}{l |l| cccccc| cccccc}
    \hline
    \multirow{2}{*}{Location} &
    \multirow{2}{*}{Distribution} &
        \multicolumn{5}{c}{LOS} & 
        \multicolumn{5}{c}{NLOS} \\ 
    & & KS Stat &  p-value & R & Loc &  Scale & Shape & KS Stat & p-value & R &  Loc &  Scale & Shape\\ \hline
     \multirow{9}{*}{Campus} & Normal  & 0.095    & \num{9e-08} & 0.96 & -25.57  & 8.68 & -  & 0.162& \num{0.80}
 & 0.94 & -25.57 & 8.68  & - \\
    & Exponential & 0.31 & \num{4e-83} & 0.97 &  -45.83 &  20.25 & - & 0.13 & 0.94 & 0.97 & -45.83  & 20.26  & -  \\
    & Log-Normal & \textbf{0.052} & \textbf{0.011} & \textbf{0.98} & -55.03 & 28.28  & 0.28 & \textbf{0.115}  & \textbf{0.98} & \textbf{0.98} & -55.03  & 28.28  & 0.28  \\
    & Rayleigh & 0.107 & \num{8e-10} & 0.98 & -37.17  & 10.24 & - & 0.14& 0.90 & 0.98 & -37.1 & 10.24  & -  \\
    & Rician & 0.101 & \num{1e-08} & 0.96 & -46.04  & 9.39  & 1.92 & 0.14& 0.90 & 0.97 & -46.04 & 9.39 & 1.92 \\
    & Nakagami  & 0.078& \num{2e-05}& 0.97& -46.57  & 22.72 &  1.66 & 0.21 & 0.50 & 0.97 & -41.57  & 22.72  & 1.66\\
    & Gamma  & \textbf{0.060}& \textbf{0.002} & \textbf{0.98} &-25.57 & 8.47  & 0.74 & 0.22 & 0.39 & 0.99 & -48.56  & 3.12  & 7.36 \\
    & Beta & \textbf{0.060}& \textbf{0.002} & \textbf{0.98} & -48.56 & 3.12 & 7.36  &0.19 & 0.58 & 0.98 & -25.57 & 8.47 & 0.73  \\
    & Log Logistic& 0.690  & 0& 0.49 &  -45.83   & 1.34  & 0.57  &0.33& 0.07 & 0.73 & -45.83 & 1.34 & 0.57  \\ \hline
    
    \multirow{9}{*}{City} & Normal  & 0.079   & \num{1e-04} & 0.97 & -26.75  & 8.14& -  & 0.112& \num{3.5e-5}
 & 0.96 & -26.09 & 6.86  & - \\
    & Exponential & 0.37 & \num{3e-93} & 0.96 &  -50.50 &  23.75 & - & 0.3 & \num{3.51 e-05} & 0.98 & -39.66  & 13.56 & -  \\
    & Log-Normal & \textbf{0.059} & \textbf{0.008} & \textbf{0.98} & -85.46  & 58.16 & 0.13 & \textbf{0.059}  & \textbf{0.087} & \textbf{0.99} & -49.29  & 22.27  & 0.28  \\
    & Rayleigh & 0.59 & \num{1e-254} & 0.97 & -27.09  & 5.76  & - & 0.099& \num{3e-04} & 0.98 & -39.85 & 10.86  & -  \\
    & Rician & 0.98 & 0 & 0.97 & -49.31  & 2.03  &1.3 & 0.099& \num{3e-4} & 0.98 & -39.86 & 10.87 & \num{5e-04}  \\
    & Nakagami  & 0.07& \textbf{\num{9e-04}} & \textbf{0.97}& -62.62  & 36.78 & 5.07 & 0.08 & \num{6e-03} & 0.98 & -40.65  & 16.09 & 1.3 \\
    & Gamma  & \textbf{0.060}& \textbf{0.002} & \textbf{0.99} &-32.0  & 3.51  & 4.3 & 0.048 & \num{3e-04} & 0.98 & -44.4  & 2.1  & 9.56 \\
    & Beta & 0.062& 0.005 & 0.98 & -26.75 & 8.04 & 0.36  &0.063 & 0.063 & 0.99 & -26.09 & 6.7 & 0.76  \\
    & Log Logistic& \textbf{0.027}  & \textbf{0.59}& \textbf{0.98} &  -78.07   & 50.64  & 12  &0.44& \num{2e-78}& \textbf{0.67} & -39.66 & 5.53 & 0.8  \\ \hline
    \multirow{9}{*}{Residential} & Normal  & 0.11    & \num{8e-06} & 0.96 & -22.05  & 8.86 & -  & 0.10& 0.045
 & 0.97 & -19.24 & 7.21  & - \\
    & Exponential & 0.29 & \num{4e-38} & 0.97 &  -40 &  17.95 & - & 0.1 &\num{1e-04} & 0.98 & -30.47  & 11.23  & -  \\
    & Log-Normal & \textbf{0.056} & \textbf{0.01} & \textbf{0.98} & -50.82 & 27.49  & 0.3 & \textbf{0.05}  & \textbf{0.69} & \textbf{0.99} & -34.64  & 13.79 & 0.47 \\
    & Rayleigh & 0.107 & \num{8e-10} & 0.98 & -30.42  & 11  & - & 0.059& \num{5e-06} & 0.98 & -31.75& 10.21  & -  \\
    & Rician & 0.111 & \num{1e-05} & 0.98 & -40.19  & 21.07 & 1.31 & 0.081& 0.165 & 0.99 & -31.75 & 10.21 & \num{4e-04}\\
    & Nakagami  & 0.08& \num{3e-03} & 0.98& -41.17  & 21.07 &  1.31& \textbf{0.04} & \textbf{0.83}& \textbf{0.99} & -30.56  & 13.43  & 0.71 \\
    & Gamma  & \textbf{0.062}& \textbf{0.045} & \textbf{0.99} &-22.05  & 8.72  & 0.78 & \textbf{0.042}& \textbf{0.877} & \textbf{0.99} & -31.13  & 4.73  & 2.51 \\
    & Beta & \textbf{0.062}& \textbf{0.045} & \textbf{0.99} & -16.7 & 7.3 & 0.95  &\textbf{0.042}  & \textbf{0.877} & \textbf{0.99} & -19.24 & 7.5 & 1.26  \\
    & Log Logistic& \textbf{0.039}  & \textbf{0.42}& \textbf{0.98} &  -46.73   & 23.13  & 5.042  &0.055& 0.605 & 0.99 & -33.02 & 12.13  & 3.1 \\ \hline
    \end{tabular}
    
\end{table*}
\section{Summary statistics of normalized delay distribution- (compatible with python scipy.stats library) }
\label{SecondAppendix}

Tables \ref{tab:NDD - Results - Summary} and \ref{tab:NDD - Results -Outdoor Summary} provided summary statistics of fitting different distributions for indoor and outdoor locations respectively. Unlike \ac{NPD}, since the lowest delay encountered from the normalized delay distribution is zero we have forced each distribution to locate at zero.
\begin{table} [t]
    \centering
    \caption{Summary statistics of the normalized delay distribution in indoor locations- (compatible with python scipy.stats library)}
    \label{tab:NDD - Results - Summary}
    \resizebox{\columnwidth}{!}{\begin{tabular}{l |l| ccccc| ccccc}
    \hline
    \multirow{2}{*}{Location} &
    \multirow{2}{*}{Distribution} &
        \multicolumn{4}{c}{LOS} & 
        \multicolumn{4}{c}{NLOS} \\ 
    & & KS Stats & p-value & R & Loc &  Scale  & KS Stats & p-value & R &  Loc &  Scale \\ \hline
   \multirow{2}{*}{Sello} &  Exponential & \textbf{0.07}&\textbf{0.04} & \textbf{0.99} &  0 &  50.52 & 0.14&\textbf{0.56} & \textbf{0.99}  & 0  & 43.51    \\
    & Weibull & 0.09& 0.01 & 0.99 & 0&  52.72 & 0.82 & \num{4e-22} & 0.91& 0  & 1.05  \\ \hline
    
    \multirow{2}{*}{Airport} & Exponential & \textbf{0.05} & \textbf{0.21} & \textbf{0.99} &  0 &  69.7  &\textbf{0.09}& \textbf{0.8} & \textbf{0.99} & 0 & 81.21 \\
    & Weibull & 0.06& 0.09 & 0.99 & 0& 71.65  &\textbf{0.97}& \num{2e-66} & 0.90 & 0   & 1.05  \\ \hline
    
    \multirow{2}{*}{TUAS}  & Exponential & \textbf{0.17}&\textbf{0.16} & \textbf{0.99} &  0 &  26.5  & 0.09&\textbf{0.002} & \textbf{0.99} & 0 & 63.2  \\
    & Weibull&0.84 &\num{4e-32} & 0.91 & 0 & 1.05&0.91  & 0 & 0.99 & 0   & 1.05   \\ \hline
    
    \multirow{2}{*}{TUAS2}& Exponential &\textbf{0.06} & \textbf{0.28} & \textbf{0.99} &  0 &  37.5 &\textbf{0.05} & \textbf{0.0001}& \textbf{0.99} & 0 & 55.7   \\
    & Weibull & 0.9 & \num{1e-262}& 0.99 & 0 & 1.05 &0.61 & 0 & 0.99 & 0 & 1.05  \\ \hline
    \end{tabular}}
\end{table}

\begin{table} [t]
    \centering
    \caption{Summary statistics of the normalized delay distribution in outdoor locations- (compatible with python scipy.stats library)}
    \label{tab:NDD - Results -Outdoor Summary}
    \resizebox{\columnwidth}{!}{\begin{tabular}{l |l| ccccc| ccccc}
    \hline
    \multirow{2}{*}{Location} &
    \multirow{2}{*}{Distribution} &
        \multicolumn{4}{c}{LOS} & 
        \multicolumn{4}{c}{NLOS} \\ 
    & &  KS Stats &p-value & R & Loc &  Scale  & KS Stats & p-value & R &  Loc &  Scale \\ \hline
   \multirow{2}{*}{Campus} &  Exponential &0.11 & \num{6e-12} & 0.99 &  0 &  136.2& \textbf{0.33}&\textbf{0.06} & \textbf{0.91} & 0  & 107.38    \\
    & Weibull &0.057 & \num{4e-4} & 0.99 & 0& 136.19& 0.05&0.004 & 0.99 & 0  & 109.93  \\ \hline
    
    \multirow{2}{*}{City}  & Exponential &\textbf{0.04}& \textbf{0.10} & \textbf{0.99}&  0 &  135.25  & 0.103 &\num{1e-4}&0.95& 0 & 81.05  \\
    & Weibull &0.05&0.02& 0.99 & 0 & 139& 0.17 & \num{1e-11} & 0.95 & 0   & 97.87  \\ \hline
    
    \multirow{2}{*}{Residential} & Exponential &0.08 &\num{2e-3}& 0.99&  0 &  124.89  & 0.21&\num{1e-07}& 0.96 & 0 & 98.05 \\
    & Weibull & 0.13 & \num{5e-08} & 0& 149.5 &0.99 & 0.21 & \num{1e-7} & 0.98&0   & 127  \\ \hline
    
    \end{tabular}}
\end{table}

\section{Summary statistics of distribution of number of paths}
\label{ThirdAppendix}

Summary statistics of KS test on fitting different distributions to empirical data for \ac{NoP} is provided in Tables \ref{tab:NoP - indoor- Results - Summary} and \ref{tab:NoP - outdoor- Results - Summary} for indoor and outdoor locations respectively.
\begin{table} [t]
    \centering
    \caption{Summary statistics of NoPs distribution in indoor locations- (compatible with python scipy.stats library)}
    \label{tab:NoP - indoor- Results - Summary}
    
    \begin{tabular}{l|l |l| ccc}
    \hline
    {Scenario}&{Location} &
    {Distribution} &
        KS Stat &  p-value  
    \\ \hline
     \multirow{18}{*}{LOS} &\multirow{9}{*}{Sello} & Normal  & 0.22   & 0.34\\
    & & Exponential & 0.28 & 0.12    \\
    & & Log-Normal & 0.15 & 0.76 \\
    & & Rayleigh & 0.19 & 0.52  \\
    & & Rician & 0.19 & 0.52   \\
    & & Nakagami  & 0.19 & 0.52   \\
    & & Gamma  & 0.167 & 0.71 \\
    & & Beta & 0.167 & 0.71 \\
    & & Log Logistic& 0.127  & 0.92 \\ 
    \cline{2-5}
    &\multirow{9}{*}{Airport} & Normal  & 0.19    & 0.79  \\
    & & Exponential & 0.19 & 0.8  \\
    & & Log-Normal & 0.45 & 0.02 \\
    & & Rayleigh & 0.19 & 0.79  \\
    & & Rician  & 0.19 & 0.79  \\
    & & Nakagami  & 0.25 & 0.44\\
    & & Gamma  & 0.26 & 0.404 \\
    & & Beta & 0.19 & 0.76 \\
    & & Log Logistic& 0.162& 0.92   \\ \hline
     \multirow{18}{*}{NLOS} &\multirow{9}{*}{TUAS} & Normal  & 0.15    & 0.67   \\
    & & Exponential & 0.15 & 0.67  \\
    & & Log-Normal & 0.42 & \num{7e-7} \\
    & & Rayleigh & 0.11 & 0.9 \\
    & & Rician & 0.11 & 0.89   \\
    & & Nakagami  & 0.19 & 0.38  \\
    & & Gamma  & 0.35 & 0.006 \\
    & & Beta & 0.24& 0.143 \\
    & & Log Logistic& 0.513  & \num{1e-05}\\ 
    \cline{2-5}
    &\multirow{9}{*}{TUAS2} & Normal  & 0.136    & 0.189 \\
    & & Exponential & 0.118 & 0.33 \\
    & & Log-Normal & 0.106 & 0.46 \\
    & & Rayleigh & 0.154 & 0.099 \\
    & & Rician & 0.154 & 0.099 \\
    & & Nakagami  & 0.072 & 0.88 \\
    & & Gamma  & 0.093& 0.62 \\
    & & Beta & 0.093& 0.62  \\
    & & Log Logistic& 0.56  & \num{8e-19} \\ \hline
    
    \end{tabular}
    
\end{table}

\begin{table} [t]
    \centering
    \caption{Summary statistics of MPCs distribution in outdoor locations- (compatible with python scipy.stats library)}
    \label{tab:NoP - outdoor- Results - Summary}
    
    \begin{tabular}{l|l |l| cc}
    \hline
    {scenario}&{Location} &
    {Distribution} &
        KS Stat &  p-value  
    \\ \hline
     \multirow{18}{*}{LOS} &\multirow{9}{*}{Campus} & Normal  & 0.18    & 0.17 \\
    & & Exponential & 0.18 & 0.17 \\
    & & Log-Normal & 0.40 & \num{1e-5} \\
    & & Rayleigh & 0.15 & 0.34\\
    & & Rician & 0.15 & 0.34   \\
    & & Nakagami  & 0.13& 0.54 \\
    & & Gamma  & 0.10& 0.8 \\
    & & Beta & 0.10& 0.8 \\
    & & Log Logistic& 0.09 & 0.86   \\ 
    \cline{2-5}
    &\multirow{9}{*}{Residential} & Normal  & 0.22   & 0.14  \\
    & & Exponential & 0.13 & 0.7  \\
    & & Log-Normal & 0.09 & 0.96  \\
    & & Rayleigh & 0.25 & 0.05  \\
    & & Rician &  0.25 & 0.05  \\
    & & Nakagami  & 0.26& 0.05\\
    & & Gamma  & 0.154 & 0.53   \\
    & & Beta & 0.11& 0.85  \\
    & & Log Logistic& 0.091 & 0.97  \\ \hline
     \multirow{18}{*}{NLOS} &\multirow{9}{*}{City} & Normal  & 0.15   & 0.53    \\
    & & Exponential & 0.181 & 0.34   \\
    & & Log-Normal & 0.625 & \num{5e-10}   \\
    & & Rayleigh & 0.139 & 0.66 \\
    & & Rician & 0.139 & 0.66  \\
    & & Nakagami  & 0.31& 0.013\\
    & & Gamma  & 0.233 & 0.11  \\
    & & Beta & 0.22 & 0.121 \\
    & & Log Logistic& 0.191  & 0.28  \\ 
    \cline{2-5}
    &\multirow{9}{*}{Residential} & Normal  & 0.17    & 0.086 \\
    & & Exponential & 0.34 & \num{2e-5}  \\
    & & Log-Normal & 0.34 & \num{2e-5} \\
    & & Rayleigh & 0.23 & 0.013  \\
    & & Rician & 0.22 & 0.013  \\
    & & Nakagami  & 0.404& \num{1e-07} \\
    & & Gamma  & 0.36& \num{5e-06}    \\
    & & Beta & 0.34 & \num{2e-05} \\
    & & Log Logistic& 0.34 & \num{2e-05} \\ \hline
    
    \end{tabular}
    
\end{table}

\clearpage
\bibliographystyle{IEEEtran}
\bibliography{references}

\end{document}